\documentclass[a4paper,nofootinbib,preprintnumbers]{revtex4}\usepackage[pdftex]{graphicx}
\usepackage{multirow}
\usepackage{color}

\newcommand{\nc}{\newcommand}
\nc{\be}[1]{\begin{equation}\mbox{$\label{#1}$}}
\nc{\bea}[1]{\begin{eqnarray} \mbox{$\label{#1}$}}
\nc{\Section}[2]{\section{#2}\label{#1}}
\nc{\Bibitem}[1]{\bibitem{#1}}
\nc{\Label}[1]{\label{#1}}

\nc{\ev}{\mathcal{E}}

\nc{\eea}{\end{eqnarray}}
\nc{\ee}{\end{equation}}

\nc{\bdm}{\begin{displaymath}}
\nc{\edm}{\end{displaymath}}
\nc{\dpsty}{\displaystyle}
\nc{\bc}{\begin{center}}
\nc{\ec}{\end{center}}
\nc{\ba}{\begin{array}}
\nc{\ea}{\end{array}}
\nc{\bab}{\begin{abstract}}
\nc{\eab}{\end{abstract}}
\nc{\btab}{\begin{tabular}}
\nc{\etab}{\end{tabular}}
\nc{\bit}{\begin{itemize}}
\nc{\eit}{\end{itemize}}
\nc{\ben}{\begin{enumerate}}
\nc{\een}{\end{enumerate}}
\nc{\bfig}{\begin{figure}}
\nc{\efig}{\end{figure}}

\nc{\arreq}{&\!=\!&}
\nc{\arrmi}{&\!-\!&}
\nc{\arrpl}{&\!+\!&}
\nc{\arrap}{&\!\!\!\approx\!\!\!&}
\nc{\non}{\nonumber}
\nc{\align}{\!\!\!\!\!\!\!\!&&}

\def\lsim{\; \raise0.3ex\hbox{$<$\kern-0.75em
      \raise-1.1ex\hbox{$\sim$}}\; }
\def\gsim{\; \raise0.3ex\hbox{$>$\kern-0.75em
      \raise-1.1ex\hbox{$\sim$}}\; }

\nc{\DOT}{\hspace{-0.08in}{\bf .}\hspace{0.1in}}
\nc{\Laada}{\hbox {$\sqcap$ \kern -1em $\sqcup$}}
\nc\loota{{\scriptstyle\sqcap\kern-0.55em\hbox{$\scriptstyle\sqcup$}}}
\nc\Loota{{\sqcap\kern-0.65em\hbox{$\sqcup$}}}
\nc\laada{\Loota}
\nc{\qed}{\hskip 3em \hbox{\BOX} \vskip 2ex}

\nc{\real}{{\rm I \! R}}
\nc{\Z}{{\sf Z \!\!\! Z}}
\nc{\complex}{{\rm C\!\!\! {\sf I}\,\,}}
\def\bigid{\leavevmode\hbox{\small1\kern-3.8pt\normalsize1}}
\def\id{\leavevmode\hbox{\small1\kern-3.3pt\normalsize1}}
\nc{\slask}{\!\!\!/}
\nc{\bis}{{\prime\prime}}
\nc{\pa}{\partial}
\nc{\na}{\nabla}
\nc{\ra}{\rangle}
\nc{\la}{\langle}
\nc{\goto}{\rightarrow}
\nc{\swap}{\leftrightarrow}

\nc{\EE}[1]{ \mbox{$\cdot10^{#1}$} }
\nc{\abs}[1]{\left|#1\right|}
\nc{\at}[2]{\left.#1\right|_{#2}}
\nc{\norm}[1]{\|#1\|}
\nc{\abscut}[2]{\Abs{#1}_{\scriptscriptstyle#2}}
\nc{\vek}[1]{{\rm\bf #1}}
\nc{\integral}[2]{\int\limits_{#1}^{#2}}
\nc{\inv}[1]{\frac{1}{#1}}
\nc{\dd}[2]{{{\partial #1}\over{\partial #2}}}
\nc{\ddd}[2]{{{{\partial}^2 #1}\over{\partial {#2}^2}}}
\nc{\dddd}[3]{{{{\partial}^2 #1}\over
    {\partial #2 \partial #3}}}
\nc{\dder}[2]{{{d #1}\over{d #2}}}
\nc{\ddder}[2]{{{d^2 #1}\over{d {#2}^2}}}
\nc{\dddder}[3]{{d^2 #1}\over
    {d #2 d #3}}
\nc{\dx}[1]{d\,^{#1}x}
\nc{\dy}[1]{d\,^{#1}y}
\nc{\dz}[1]{d\,^{#1}z}
\nc{\dl}[1]{\frac{d\,^{#1}l}{(2\pi)^{#1}}}
\nc{\dk}[1]{\frac{d\,^{#1}k}{(2\pi)^{#1}}}
\nc{\dq}[1]{\frac{d\,^{#1}q}{(2\pi)^{#1}}}

\nc{\bfT}{{\bf T }}

\nc{\cA}{{\cal A}}
\nc{\cB}{{\cal B}}
\nc{\cD}{{\cal D}}
\nc{\cE}{{\cal E}}
\nc{\cG}{{\cal G}}
\nc{\cH}{{\cal H}}
\nc{\cL}{{\cal L}}
\nc{\cO}{{\cal O}}
\nc{\cT}{{\cal T}}
\nc{\cN}{{\cal N}}
\nc{\cR}{{\cal R}}
\nc{\rvac}[1]{|{\cal O}#1\rangle}
\nc{\lvac}[1]{\langle{\cal O}#1|}
\nc{\rvacb}[1]{|{\cal O}_\beta #1\rangle}
\nc{\lvacb}[1]{\langle{\cal O}_\beta #1 |}
\nc{\bb}{\bar{\beta}}
\nc{\bt}{\tilde{\beta}}
\nc{\ctH}{\tilde{\cal H}}
\nc{\chH}{\hat{\cal H}}
\nc{\1}{\aa}
\nc{\2}{\"{a}}
\nc{\3}{\"{o}}
\nc{\4}{\AA}
\nc{\5}{\"{A}}
\nc{\6}{\"{O}}
\nc{\al}{\alpha}
\nc{\g}{\gamma}
\nc{\Del}{\Delta}
\nc{\e}{\textrm{e}}
\nc{\eps}{\epsilon}
\nc{\lam}{\lambda}
\nc{\Om}{\Omega}
\nc{\ve}{\varepsilon}
\nc{\mn}{{\mu\nu}}
\nc{\vp}{\varphi}

\nc{\rf}[1]{(\ref{#1})}
\nc{\nn}{\nonumber \\*}
\nc{\bfB}{\bf{B}}
\nc{\bfv}{\bf{v}}
\nc{\bfx}{\bf{x}}
\nc{\bfy}{\bf{y}}
\nc{\vx}{\vec{x}}
\nc{\vy}{\vec{y}}
\nc{\oB}{\overline{B}}
\nc{\oI}{\overline{I}}
\nc{\oR}{\overline{R}}
\nc{\rar}{\rightarrow}
\nc{\ti}{\times}
\nc{\slsh}{\hskip-5pt/}
\nc{\sm}{Standard~Model~}
\nc{\MP}{M_{\rm Pl}}
\nc{\mpl}{M_{\rm Pl}}
\nc{\tp}{t_{\rm Pl}}

\nc{\pmin}{p_{\rm min}}
\nc{\pmax}{p_{\rm max}}
\nc{\fo}{f_0}
\nc{\foi}{f_{0,i}\,}
\nc{\fop}{f_0^P}
\nc{\fou}{f_0^U}

\nc{\eff}{{\rm eff}}
\nc{\MT}{M_{\rm T}}
\nc{\ML}{M_{\rm L}}
\nc{\kk}{\vek{k}}
\nc{\pp}{{\rm p}}
\nc{\pt}{\partial_t}
\nc{\half}{{1\over 2}}
\nc{\w}{\omega}
\nc{\uhat}{\hat{U}_\w}

\nc{\etal}{\mbox{\it et al.}}
\nc{\ie}{{\it i.e. }}
\nc{\eg}{{\it e.g. }}
\nc{\trh}{T_{\rm RH}}
\nc{\ad}{{a'\over a}}
\nc{\bd}{{b'\over b}}
\nc{\Rd}{{R'\over R}}
\nc{\diag}{{\textrm{diag}}}
\nc{\mato}[1]{\tilde{#1}}
\nc{\sech}{\textrm{sech}}
\nc{\I}{\textrm{I}}
\nc{\II}{\textrm{II}}
\nc{\III}{\textrm{III}}
\nc{\vev}[1]{\langle #1 \rangle}
\nc{\hyp}{\,\; F_{1{\hskip -16pt}2}{\hskip 11pt}}
\nc{\brhom}{\overline{\rho}_M}
\nc{\brho}{\overline{\rho}}
\nc{\rhob}{\overline{\rho}}
\nc{\Pb}{\overline{P}}
\nc{\bH}{\overline{H}}
\nc{\ep}{{1+4\eps}}

\nc{\lcdm}{$\Lambda$CDM }
\nc{\ms}{\langle\sigma\rangle}
\nc{\V}{\mathcal{V}}
\nc{\Od}{\Omega_d}
\nc{\Or}{\Omega_r}
\nc{\OL}{\Omega_{\Lambda}}

\def\smiley{\hbox{\large$\bigcirc$\hspace{-.80em}%
\raise.2ex\hbox{$\cdot\cdot$}\kern-.61em    
\lower.2ex\hbox{\scriptsize$\smile$}}\ }

\def\frowney{\hbox{\large$\bigcirc$\hspace{-.80em}%
\raise.2ex\hbox{$\cdot\cdot$}\kern-.635em
\lower.2ex\hbox{\scriptsize$\frown$}}\ }

\begin{document}

\title{\bf Anisotropic cosmology and  inflation from tilted Bianchi IX model}
\author{P.~Sundell${}^{1,2}$}
\author{T.~Koivisto${}^2$}
\affiliation{${}^{1}$Turku Center for Quantum Physics, Department of Physics and Astronomy,
University of Turku, FIN-20014 Turku, Finland}
\affiliation{${}^{2}$Nordita, KTH Royal Institute of Technology and Stockholm University 
Roslagstullsbacken 23, SE-10691 Stockholm, Sweden}
\date{\today}

\begin{abstract}

The dynamics of the tilted  axisymmetric  Bianchi IX cosmological models are explored allowing energy flux in the source fluid. 
The Einstein equations and the continuity equation are presented treating the equation of state $w$ and the tilt angle of the fluid $\lambda$ as time dependent functions, but when analysing the phase space $w$ and $\lambda$ are considered free parameters and the shear, the vorticity and the curvature of the spacetime span a three-dimensional phase space that contains seven fixed points. 
One of them is an attractor that inflates the universe anisotropically, thus providing a counter example to the cosmic no-hair conjecture.
Also,  examples of a realistic though  fine-tuned cosmologies are presented wherein the rotation can grow significant towards the present epoch but the shear stays within the observational bounds. The examples suggest that the model used here can explain the parity-violating anomalies  of the cosmic microwave background. The result  significantly differs from an earlier study, where a non-axisymmetric Bianchi IX type with a tilted perfect dust source was found to induce too much shear for observationally significant vorticity.


\end{abstract}

\preprint{NORDITA-2015-70}

\maketitle

\section{Introduction}
\label{introduction}

Cosmological observations provide compelling evidence that our Universe at large scales is well described by the homogeneous, isotropic and spatially flat standard model called the \lcdm \cite{Ade:2013zuv}. The common explanation for this featurelessness is that inflation, while also at the same time giving birth to small initial seeds of structure, wipes out all inhomogeneities from the exponentially expanding background. Furthermore, even the initial structure generated by inflation seems to be of a vanilla nature, as the data remains consistent with primordial spectrum from quantum fluctuations of a single field with perfectly gaussian and isotropic statistical properties \cite{Ade:2015lrj}.  

However, not all the observational data comply with this picture, in particular the anomalies in the cosmic microwave background (CMB) \cite{Eriksen:2003db,Finelli:2011zs,Akrami:2014eta}. If of cosmological origin, these anisotropic statistical features in the temperature fluctuations of the CMB could be regarded as a hint of physics beyond the simplest \lcdm parameterisation of cosmology. As the anomalies are most significant at the largest CMB angles \cite{Flender:2013jja}, thus (roughly) corresponding to the scale of dark energy, it is natural to consider (slightly) imperfect fields in the present universe that could both accelerate the background expansion and generate the observed deviations from isotropy \cite{Koivisto:2007bp,Koivisto:2008ig,Perivolaropoulos:2014lua}, see also e.g. \cite{Battye:2009ze,Akarsu:2009gz,Sharif:2010hy,Koivisto:2010dr,Appleby:2012as}. On the other hand, the anisotropies could have been generated (or retained) by non-standard inflationary dynamics due to for example vector fields \cite{Koivisto:2008xf, Watanabe:2009ct,Hervik:2011xm,Yamamoto:2012tq,Maleknejad:2012fw,Jimenez:2014rna,Koivisto:2014gia} or quadratic curvature corrections to gravity \cite{Barrow:2005qv,Barrow:2009gx}.

However, such an anisotropic expansion during either the early (inflation) or the late (dark energy) accelerated stages of the universe, has a limited promise of actually explaining the CMB anomalies, because the latter are mainly parity-violating but shear does not distinguish handedness. It is this oddness that appears the most unexpected and difficult to explain with the existing models. Proposed late-time origin for a parity violation include spontaneous isotropy breaking \cite{Gordon:2005ai}, imperfect dark energy field \cite{Axelsson:2011gt} and effects due to our peculiar velocity with respect to the CMB \cite{Notari:2013iva}. The primordial spectrum itself could contain dipole and other odd contributions only in the context of noncommutative theories \cite{Koivisto:2010fk}, but parity-violating primordial modulations of the perturbations and thus of the resulting CMB temperature field might be generated by nontrivial multifield dynamics during inflation \cite{Mazumdar:2013yta} or reheating \cite{McDonald:2013qca}, or by breaking translation invariance with isocurvature perturbations \cite{Erickcek:2009at} or domain walls \cite{Wang:2011pb,Jazayeri:2014nya}.

The question we shall pursue here is whether our universe could be realistically described by a spacetime with slight anisotropies that distinguish orientation in way that would explain the parity-violating CMB anomalies. A natural framework for this is provided by the Bianchi class B models that describe homogeneous but anisotropic spacetimes that allow for rotation \cite{Ellis,Ellis:1998ct}. Such are sufficiently general to incorporate parity-violating cosmological features (unlike Bianchi class A models that allow only for even asymmetries) but due to remaining symmetries (we consider locally rotationally symmetric Bianchi models) simple enough to be tractable. 
Indeed, it is has been shown that the CMB anomalies can be very well fitted by a Bianchi class B type VII${}_{\rm h}$ template \cite{Eriksen:2003db,Ade:2013vbw}, but only the anomalies: the cosmological parameters of the best-fit Bianchi template model are in strong disagreement with the rest of the cosmological data \cite{McEwen:2013cka,Ade:2013vbw}. This suggests it may be worthwhile to explore 
different or more elaborate Bianchi type B cosmologies. 

We are thus lead to consider generalisations of the above cited anisotropic inflation and imperfect dark energy models that have been previously discussed, as far as we know, only in the context of Bianchi class A spacetimes. 

A rather generic property of perfect-fluid cosmological spacetimes is that they tend to isotropise, which is also the reason for the difficulty of reconciling the Bianchi VII${}_{\rm h}$ rotating universe with realistic cosmology. This property is reflected in a cosmic no-hair conjecture suggesting that anisotropies will not survive in inflating spacetimes \cite{Wald:1983ky,Rothman198619}. Behind the conjecture is a theorem stating that initially expanding homogeneous cosmological models, which satisfy Einstein's equations with positive cosmological constant and dominant and strong energy conditions, can evolve only so that they either collapse or approach the de Sitter solution at the same time making the universe isotropic. The only exception is Bianchi IX type universe, where the curvature  can come with the positive sign and could thus in principle cancel the driving effect of shear and energy density to the expansion \cite{Wald:1983ky,Rothman198619}. As mentioned above however, several examples have been found where anisotropies can be supported in the presence of imperfect fluids, and refinements of the cosmic no-hair theorem have been formulated \cite{Maleknejad:2012as}.

Motivated by these considerations, we will specialise here to cosmologies in the Bianchi class B type IX models. In order to possibly generate vorticity, we include a ''tilted'' energy source, i.e. a fluid whose rest frame does not coincide with the cosmic rest frame, and furthermore it turns out that this requires that the fluid has an energy flux that is fixed by consistency conditions. When analysing the phase space, for simplicity, we will assume that the fluid can be characterised by a constant equation of state and that the tilt angle remains constant during the cosmic evolution. This defines our two-parameter phenomenological model of tilted dark energy in a universe allowing rotation. From this simple starting point emerge already interesting possibilities such as anisotropic inflation. Since Misner's classic mixmaster paper \cite{Misner:1969hg}, studies of Bianchi IX cosmologies in the literature have been undertaken in the contexts of loop quantum cosmology \cite{Bojowald:2004ra,Battisti:2009kp}, chaotic dynamics \cite{DeOliveira:2002ih,Kim:2013xu} and viscous fluids \cite{Rao:2012cp,Kohli:2013tpa}, but not to our knowledge in the presence of tilt and heat flux. 

The structure of the paper is as follows. We first specify the model in section \ref{model} and derive its governing differential equations, then analyse its phase space in section \ref{Fixedpoints}, in particular locate the fixed points there and investigate their stability properties. In section \ref{cosmo} we then consider the possible cosmological applications of the model, focusing on the new anisotropic features we could generate without running into conflict with observational upper bounds on them. The paper is concluded by section \ref{conclu} and a few details are confined to appendices \ref{energyconditions}-\ref{decomposition}.

\section{The model} 
\label{model}

We shall work in the framework of the axisymmetric Bianchi IX cosmological model, where the line element is \cite{Misner:1969hg}
\begin{equation}\label{le}
ds^2=-dt^2+ e^{2\alpha (t)+2\beta (t)}\left[ (\omega^1)^2+(\omega^2)^2 \right]+e^{2\alpha (t)-4\beta (t)}(\omega^3)^2\,,
\end{equation}
with the one-forms $\omega^i$ given as
\begin{eqnarray} \label{oneforms}
\omega^1&=&\sin \psi d\theta-\sin \theta \cos \psi d \phi\,, \\
\omega^2&=&\cos \psi d\theta+\sin \theta \sin \psi d \phi\,, \\
\omega^3&=&\cos \theta d \phi + d \psi\,.
\end{eqnarray}
There are 5 non-zero components of the Einstein tensor $E^a_b$ using the line element (\ref{le}), however,  $E^1_1=E^2_2$ and $E^2_3=\cos(\theta)(E^3_3-E^1_1)$, thus only three of them are independent. 

We consider the energy-momentum tensor to be composed from perfect fluid and energy flux,
\begin{equation}\label{Tab2}
T^a_b=[\rho(t)+p(t)]u^au_b+p(t)g^a_b+q^a u_b+u^aq_b\,,
\end{equation}
where $\rho(t)$ is the  matter density, $p(t)$ is the pressure, and the energy flux vector $q_a$ is orthogonal to $u_a$. We take the four-velocity to be
\begin{eqnarray}\label{u}
u^a&=&\left(\cosh [\lambda(t)],0,0,\sinh [\lambda(t)]
   e^{2 \beta (t)-\alpha (t)}\right)\,, 
\end{eqnarray}  
which satisfies the normalisation condition  $u^au_a=-1$. The ''tilt'' vanishes in the limit $\lambda \rightarrow 0$. The fluid we assume to obey,
\be{es}
p(t)= w(t) \rho(t) \,.
\ee

Using the non-diagonal components of  the Einstein equations\footnote{We set $8\pi G = 1$.} ($E^a_b=T^a_b$) and the orthogonality condition ($q_a u^a=0$), the components of the energy flux vector are found:
\be{q}
q_0=-q_3 \tanh [\lambda(t)]e^{2 \beta (t)-\alpha (t)}, \quad q_1=0, \quad q_2=q_3 \cos \theta, \quad q_3=- (p(t)+\rho(t) ) \frac{ \sinh [\lambda(t)] \cosh ^2 [\lambda(t)]}{ \text{cosh}  [2\lambda(t)] }e^{\alpha (t)-2 \beta (t)}.
\ee 
Hence, after the dynamics of $\alpha$, $\beta$, $p$, $\rho$, $\lambda$, and $\theta$ are known,  the dynamics of the energy flux is describable. Therefore, we can study the dynamics of the system without paying attention to the energy flux and consequently we neglect it from now on. However, the presence of the energy flux in the energy-momentum tensor is crucial, because in the absence of the energy flux the equations would be consistent only if $w=-1$ or $\lambda=0$. In a sense, the role of the heat flux is to allow $w$ and $\lambda$ vary freely. The energy conditions for the energy-momentum tensor (\ref{Tab2}) are not affected by the tilt and heat flux as discussed in the Appendix \ref{energyconditions}. Moreover, by defining an auxiliary function
\be{W}
W(t) \equiv \frac{1}{2}\left( 1-w(t)+\frac{w(t)+1}{ \text{cosh} [2\lambda(t)]} \right),
\ee
 the energy-momentum tensor (\ref{Tab2}) can now be recast as
\be{emtuus}
T=\left(
\begin{array}{cccc}
 -W(t) \rho(t)  & 0 & 0 & 0 \\
 0 & w(t) \rho(t)  & 0 & 0 \\
 0 & 0 & w(t) \rho(t)  & 0 \\
 0 & 0 & [W(t)-1] \rho(t)  \cos (\theta ) & [w(t)+W(t)-1] \rho(t) 
\end{array}
\right).
\ee

We can now determine the suitable kinematical scalars and spatial curvature of the metric, in terms of which to present the Einstein field equations. The covariant definitions of the kinematical quantities in the 3+1 formalism read as follows \cite{Ehlers:1993gf,Ellis:1998ct} (where the covariant time and space derivatives are defined, respectively, as $\dot{f}=u^a f_{,a}$ and $f_{;a}=(\delta^b_a+u^b u_a)f_{,b}$):
\be{qdefs}
\omega_{ab}\equiv u_{[a;b]}-\dot{u}_{[a}u_{b]}\,, \quad \sigma_{ab}\equiv u_{(a;b)}-\dot{u}_{(a}u_{b)}-\frac{1}{3} \Theta h_{ab}\,, \quad \Theta \equiv u^a_{\, \, \, ;a}, \quad \dot{u}^a \equiv u^a_{\, \, \, ;b}u^b\,. 
\ee
These are, respectively, the vorticity, the traceless shear, the expansion and the acceleration tensors associated to the fluid and subject to the conditions $\omega_{(ab)}= \sigma_{[ab]}=0$, $\omega_{ab}u^b= \sigma_{ab}u^b=0$, $\sigma^a_a=0$ and $u_a\dot{u}^a=0$. From these quantities we can then construct the kinematical scalars, that for the line element (\ref{le}) and the  four-velocity (\ref{u}) turn out to as 
\begin{eqnarray}\label{kq1}
 \omega&  \equiv & \left(\omega_{ab} \omega^{ab}/2\right)^{1/2}=\frac{1}{2} \sinh[\lambda(t)] e^{- (\alpha(t)+4 \beta (t))}\,, \nonumber  \\
\sigma & \equiv& \left(\sigma_{ab} \sigma^{ab}/2\right)^{1/2} =\sqrt{3}\cosh[\lambda(t)] \beta'(t)\,, \\
\dot{u} &\equiv & (\dot{u}_a\dot{u}^a)^{1/2}=\sinh[\lambda(t)][\alpha '(t)-2\beta '(t)]\,, \nonumber \\
\Theta&=&3\cosh[\lambda(t)] \alpha'(t) \nonumber\,,
\end{eqnarray}
where  $\omega$ is the  vorticity, $\sigma$ is the  shear, $\Theta$ is the  expansion, and  $\dot{u}$  is the acceleration scalar, respectively. The spatial curvature scalar (i.e. the Ricci scalar of the 3-dimensional constant-$t$ hypersurfaces) for the line element (\ref{le}) is
\begin{equation}\label{3R}
{}^3R=\frac{1}{2} \left(4 e^{6 \beta (t)}-1\right) e^{-2 (\alpha
   (t)+4 \beta (t))}\,.
\end{equation} 
Using Eqs.\  (\ref{u}), (\ref{es}) and (\ref{emtuus}) and the above definitions, the  Einstein equations $E^a_b=T^a_b$ assume the form of a constraint equation and two dynamical first-order differential equations:
\bea{EEs}
 -W(t) \rho(t) &=&-\frac{{}^3R(t)}{2}-\frac{\Theta(t)^2}{3 \cosh^2 [\lambda(t)]}+\frac{\sigma(t)^2}{\cosh^2 [\lambda(t)]}\,, \nonumber \\
w(t) \rho(t)&=& -\frac{\omega(t)^2}{\sinh^2 [\lambda(t)]}-\frac{\Theta(t)^2}{3 \cosh^2 [\lambda(t)]}+\frac{\Theta(t) \sigma(t)}{\sqrt{3} \cosh^2 [\lambda(t)]}-\frac{\sigma(t)^2}{\cosh^2 [\lambda(t)]}-\frac{2\Theta'(t)}{3 \cosh [\lambda(t)]}+\frac{\sigma'(t)}{\sqrt{3}\cosh [\lambda(t)]} \nonumber \\
\left[ w(t)+W(t)-1 \right]\rho(t) &=&-\frac{{}^3R(t)}{2} +\frac{2\omega(t)^2}{\sinh^2 [\lambda(t)]}-\frac{\Theta(t)^2}{3 \cosh^2 [\lambda(t)]}-\frac{2\Theta(t) \sigma(t)}{\sqrt{3} \cosh^2 [\lambda(t)]}  \\
&&-\frac{\sigma(t)^2}{\cosh^2 [\lambda(t)]}-\frac{2\Theta'(t)}{3 \cosh [\lambda(t)]}-\frac{2\sigma'(t)}{\sqrt{3}\cosh [\lambda(t)]}. \nonumber
\eea
The conservation of stress energy $\nabla_a T^{ab}=0$ results in the differential equation 
\be{cont}
0=\frac{1}{3} \left(\rho (t) \text{sech} [\lambda(t)] \left(\Theta(t) (3 w(t)+4
   W(t)-1)-2 \sqrt{3} \Sigma(t) (W(t)-1)\right)+3 \left(\rho (t) W'(t)+W(t)
   \rho '(t)\right)\right)\,.
\ee 
 However, we still need to determine the evolution of ${}^3R$ and $\omega$ in order to form an autonomous system. This can be achieved by differentiating both sides of the definitions of $\omega$ and ${}^3R$ given in Eqs.\ (\ref{kq1})-(\ref{3R}), and then reapplying Eqs.\ (\ref{kq1})-(\ref{3R}), giving
\bea{d3Rdt}
{}^3R'(t)&=&-\,{}^3R(t)\left[\frac{2\Theta(t)}{3\cosh [\lambda(t)]}+\frac{2\sigma(t)}{\sqrt{3}\cosh [\lambda(t)]}\right]+\frac{4\sqrt{3}\, \omega(t)^2 \sigma(t)}{\sinh [\lambda(t)]^2\cosh [\lambda(t)]}\,, \\ \label{domegadt}
\omega'(t)&=&- \omega(t) \left(\frac{\Theta(t)}{3\cosh [\lambda(t)]}+4 \frac{\sigma(t)}{\sqrt{3}\cosh [\lambda(t)]}\right)\,.
\eea
This completes the system of dynamical equations in terms of the variables in the 1+3 formalism \cite{Ehlers:1993gf,Ellis:1998ct}.

To apply the techniques of dynamical system analysis in cosmology\footnote{For more recent applications to anisotropic cosmologies, see for example the references \cite{Koivisto:2007bp,Koivisto:2008ig,Koivisto:2008xf,Hervik:2011xm,Kim:2013xu,Koivisto:2014gia} in the introduction.} \cite{Ellis}, it is convenient to rewrite the system in terms of dimensionless, expansion-normalised variables. For this purpose we find it most suitable to employ the Hubble rate that in the presence of tilt is defined by
\begin{equation}
H(t) \equiv \frac{\Theta(t)}{3\cosh [\lambda(t)]}\,.
\end{equation}
The dynamical variables we then define as follows
\be{dqs}
\Sigma(t) \equiv \frac{\sigma(t)}{\sqrt{3}\cosh [\lambda(t)]H(t)}\,, \quad K(t)=-\frac{{}^3R(t)}{6H(t)^2}\,, \quad \V(t)=\frac{\omega(t)}{\sinh [\lambda(t)] H(t)} \,, \quad \Omega(t)=\frac{  \rho(t)}{3H(t)^2}\,.
\ee
These are identified, respectively, as the dimensionless expansion-normalised shear,  spatial curvature and vorticity associated to the spacetime geometry, and finally the energy density assoaciated to the fluid source. By assuming the universe to be monotonically expanding and treating $\alpha$ as the time variable, we have for any function $f(t)$ that
\begin{equation}\label{time}
\frac{f'(t)}{H(t)}=\frac{f'(t)}{\alpha'(t)}=f'(\alpha)\,.
\end{equation}
This choice of time variable enables us to eliminate the Hubble rate from the resulting system of equations. 
Indeed, after some simple manipulations, Eqs. (\ref{EEs}), (\ref{cont}), (\ref{d3Rdt}), and (\ref{domegadt}) can be written using the dimensionless quantities (\ref{dqs}) as
\be{S}
\Sigma'(\alpha)  =  -1+2
  K(\alpha)+\Sigma(\alpha){}^2+(\epsilon-3)
  \Sigma(\alpha)+  \V(\alpha){}^2+ \Omega (\alpha)\,, 
\ee
\be{O}
\Omega '(\alpha)
  =  \Omega(\alpha) \left(2 \epsilon +2 \Sigma (\alpha)-4 +\frac{1-3 w-2 \Sigma (\alpha)}{W(\alpha)}\right)- \frac{ \Omega(\alpha)}{W(\alpha)}W'(\alpha) \,,
\ee
\be{V}
\V'(\alpha)  =  \V(\alpha)
  \left(\epsilon-4 \Sigma(\alpha)-1\right)\,,
\ee
\be{K}
K'(\alpha)  =  -2\left\{K(\alpha)[-\epsilon+1+\Sigma(\alpha)]+\Sigma(\alpha) \V(\alpha)^2 \right\}\,,
\ee
where we have defined  $\epsilon \equiv -H'(t)/H(t)^2$. The four dynamical equations are subject to the constraints 
\be{e}
\epsilon = \frac{1}{2}  \left[ 3w  -1  \right] \Omega(\alpha)-K(\alpha) +2+\Sigma^2(\alpha)\,,  
\ee
\be{F}
1  =  W(\alpha) \Omega(\alpha)+
K(\alpha)+ \Sigma^2(\alpha)\,.
\ee
The first constraint means simply that the $\epsilon$ in the four dynamical equations should be considered as a short-hand notation for the function of the four variables (the physical meaning of $\epsilon$ is explained shortly), but the second equation, representing the Friedmann constraint, reduces the dimensionality of the system to three.

Having now set up the dynamical system in a convenient form, we proceed to analyse its phase space.

\section{The phase space analysis} \label{Fixedpoints}

In this section, $w$ and $\lambda$ are treated as variables for simplicity and we investigate the phase space of  the three-dimensional autonomous system\footnote{We could of course write down the three dynamical equations (\ref{S},\ref{O},\ref{V},\ref{K}) with the constraints (\ref{e},\ref{F}) inserted explicitly but that would hardly give more insight to the properties of the system.} obtained by eliminating the curvature using the generalised Friedmann equation (\ref{F}), employing the analysis methods described in many text books, e.g. \cite{Wiggings,Ellis}. Even though the curvature is no longer among the set of differential equations, it can be solved using the generalised Friedmann equation. We will do this in the subsequent to track the curvature.

We also follow $\epsilon$ at each fixed point. That is of interest because this so called slow-roll parameter gives us directly information about the overall expansion rate of the universe. Namely,
the geometrical mean, $a$, of the scale factor is $a=e^{\alpha}$, and hence,  
$$
\epsilon\equiv -\frac{H'(t)}{H(t)^2}=-\frac{\alpha''}{(\alpha')^2}=-\frac{a''}{a \cdot (\alpha')^2}+1\,,
$$
and thus $\epsilon<1$ indicates acceleration and $\epsilon>1$ deceleration. From Eqs. (\ref{e}) and (\ref{F}) we furthermore see that
\bea{epsilon2}
\epsilon&=&1 \qquad \qquad \qquad \qquad  \text{in the curvature filled universe}\,, \nonumber \\
\epsilon&=&3  \qquad \qquad \qquad \qquad  \text{in the shear  filled universe}\,, \nonumber \\
\epsilon&=&2-\frac{1}{1+\sech(2\lambda)} \quad \, \text{in the dust  filled universe}\,, \nonumber \\
\epsilon&=&2 \qquad \qquad \qquad \qquad  \text{in the  radiation filled universe}\,, \nonumber \\
\epsilon&=&0 \qquad \qquad \qquad \qquad   \text{in the dark energy filled universe} \nonumber \,.
\eea
At each fixed point in the phase space, $\epsilon$ is a constant.

We note that the tilt appears in the dynamical equations only in the combination $W = (-w+1+\frac{w+1}{ \text{cosh}(2 \lambda )})/2$. In the limit of vanishing tilt, $W$ always reduces to unity. In the limit of strongly tilted cosmology, $W$ reduces to $W \rightarrow 1/2$ for dust and $W\rightarrow 1/3$ for radiation. If $w=-1$, the dynamical equations are explicitly independent on $\lambda$, but the equation for the vorticity, (\ref{V}), still holds.  Thus, to possibly rotate the universe we  need only a tilted fluid. The implicit dependence on $\lambda$ is that equation (\ref{V}) does not exist iff $\lambda=0$.

After these preliminary considerations, we are ready to list our results for the fixed points and their properties. 

\subsection{The fixed points of the system}
  
We found seven solutions to the set of equations $\V'=\Omega'=\Sigma'=0$. In the following, we will present these fixed points and study their stability in the parameter space $(\lambda,w)$, where $-\infty<\lambda<\infty$ and $-1\leq w \leq 1$.  \footnote{The governing equations depend on $\lambda$ only via $\cosh (2\lambda)$. Due to the propeties of the hyperbolic cosine function, the studied range $-\infty<\lambda<\infty$ can be presented at the  range $0\leq \lambda <\infty$.} We use an asterisk on quantities evaluated at each respective fixed point.

\begin{figure}[h]
\centering
\includegraphics[scale=1]{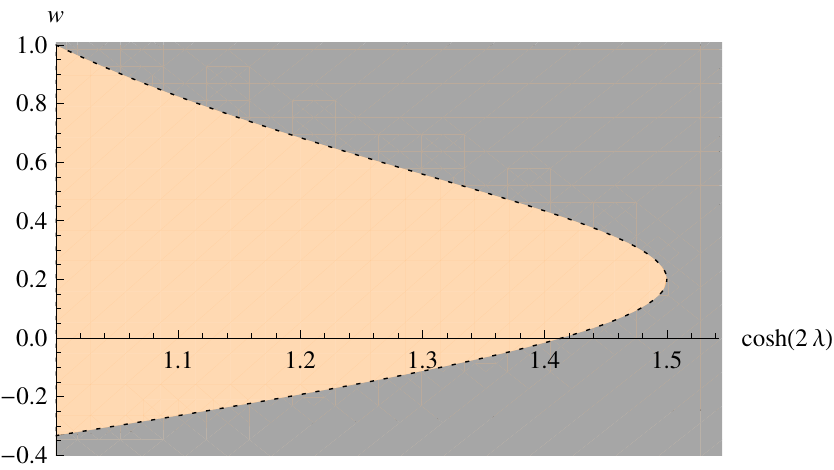} 
\includegraphics[scale=0.9]{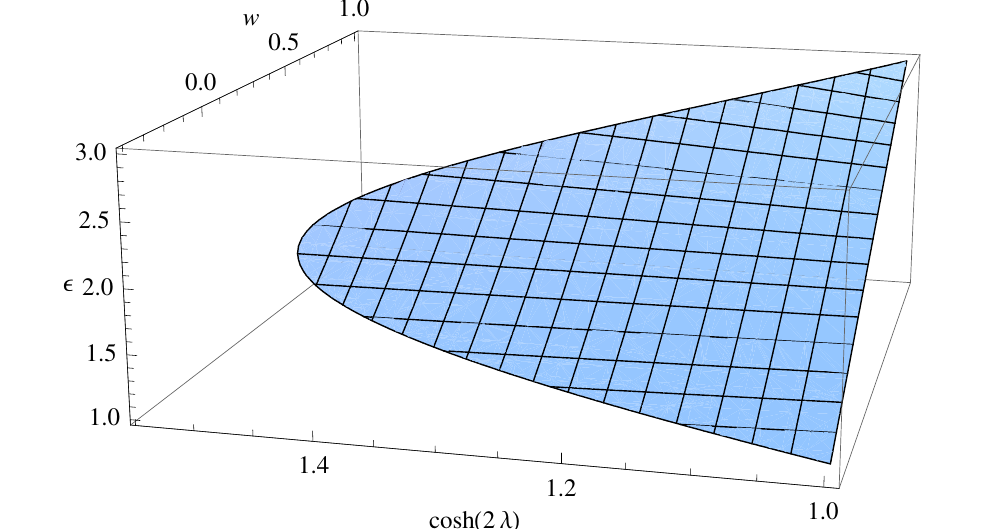} 
\caption{\footnotesize The left panel shows the stability properties of the fixed point   (\ref{fp10}): the orange area denotes saddle fixed points (repulsive in one direction) in the parameter space, the dotted curves are where the stability of the  fixed points can not be determined using the linearisation method and in the gray zone $\V$ takes imaginary values. The vorticity and the curvature are zero on the curve between the orange and the gray areas.   The right panel shows the corresponding values of $\epsilon$. Colors available online. }
\label{kfp10}
\end{figure}

\subsubsection{The fixed points allowing rotation}

The two fixed points,
\bea{fp10}
\V^*&=&\pm\frac{3}{2} \sqrt{\frac{(2-3 w) w+8 (W-1) W+1}{(1-5 W)^2}} \nonumber \\
K^*&=&\frac{3 ((2-3 w) w+8 (W-1) W+1)}{4 (1-5 W)^2} \\
\Omega^*&=& -\frac{3 (w-6 W+1)}{(1-5 W)^2} \nonumber \\
\Sigma^*&=&\frac{3 w+2 W-1}{10 W-2} \nonumber 
\eea
differing only by the sign of the vorticity $\V^*$, exist when the  inequality
\be{fp10c}
0\leq(2-3 w) w+8 (W-1) W+1
\ee
is satisfied. It is convenient to note that condition (\ref{fp10c}) implies that $0< 5 W-1$ must hold too. The  area restrained by  (\ref{fp10c}) is depicted in Figure \ref{kfp10}, where also the stability of the fixed points  is indicated by colour.  The left panel of  Figure \ref{kfp10} is drawn by evaluating the real part of each eigenvalue at each point separately. The eigenvalues are given in  Appendix \ref{Aev}. The dotted curves represent the parameter values for which at least one   eigenvalue of the stability matrix  is zero and the stability can not be determined using the linearisation method. Moreover, $K^*=\V^*=0$ on the dotted curves (for further details, see Appendix \ref{Aev1}).  The right panel of Figure \ref{kfp10} indicates  that these fixed points can represent the overall expansion of a curvature, shear, dust, or radiation dominated epochs.  

\begin{figure}[h]
\includegraphics[scale=1.05]{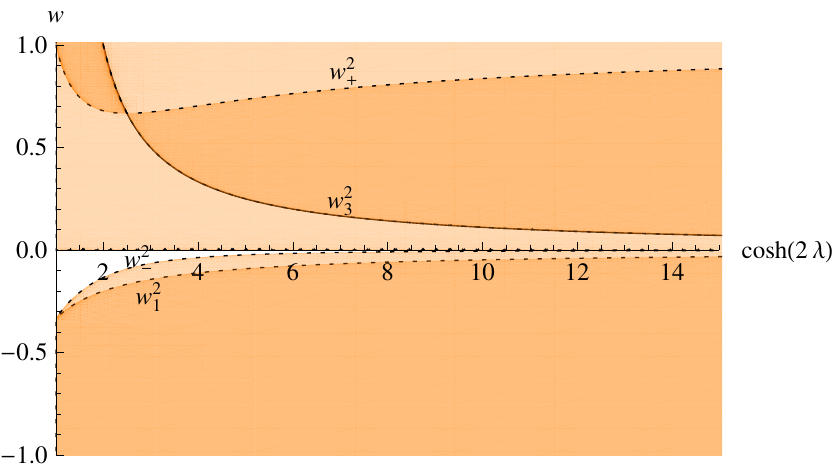}  \, \,
\includegraphics[scale=0.8]{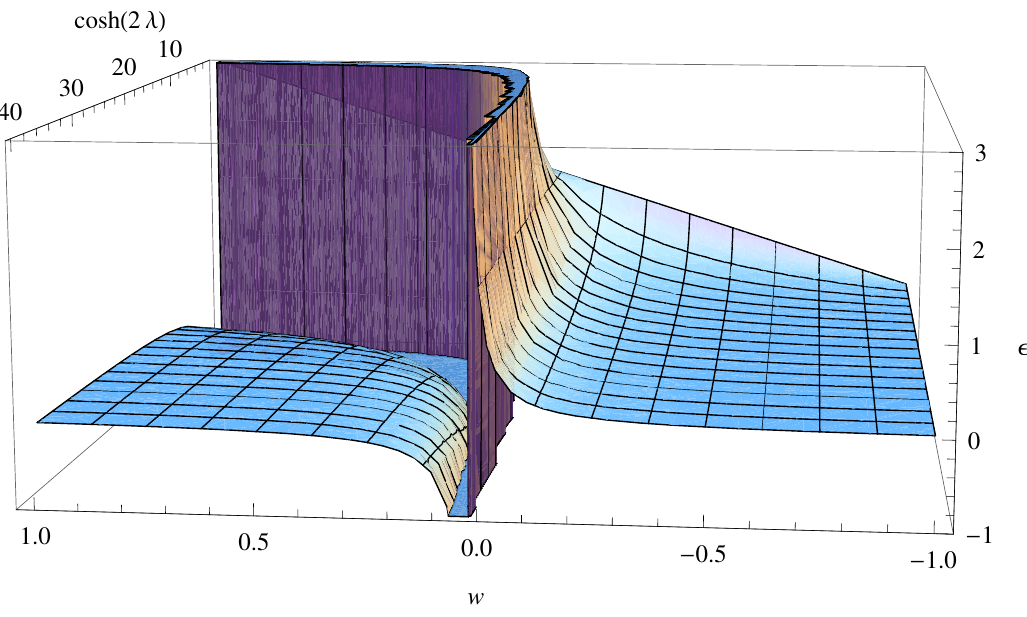} 
\caption{\footnotesize  The left panel shows the stability properties of the fixed point (\ref{fp9b}). In the orange areas the solution is a saddle point (repulsive in one directions in the lighter colored regions and in two directions in the darker colored regions), in the white area the fixed point is stable, 
and on the black curves the fixed point does not exist. Dotted curves (including $w=0$) denote where the linearization method can not recognise the stability properties. If $\lambda \rightarrow \infty$, then $w_+^2 \rightarrow 1$ and $w_-^2$, $w_1^2$, and $w_3^2$ converges to zero. The right panel shows the corresponding  $\epsilon$-values. Colors available online. }
\label{kfp98b}
\end{figure}

\subsubsection{The fixed point where only the vorticity does not exist}

The fixed point where all the effective energy sources contribute to the expansion (but the vorticity disappears) is 
\bea{fp9b}
V^*&=&0 \nonumber \\
K^*&=&\frac{3}{4} \left(\frac{(2-3 w) w}{(1-2 W)^2}+1\right) \\
\Omega^*&=&-\frac{3 w}{(1-2 W)^2} \nonumber \\
\Sigma^*&=&\frac{3 w+2 W-1}{4 W-2} \nonumber
\eea
The stabilities and values of the slow-roll parameter $\epsilon$ are depicted in the parameter space in Figure \ref{kfp98b}. 
 These fixed points can present any expansion rate as $\epsilon$ can take the value of any real number. The curve where the fixed points are not defined is  $w=[\cosh(2\lambda)-1]^{-1}$, furthermore  $\epsilon \rightarrow \infty$ when approaching the curve from the negative $w$ side and $\epsilon \rightarrow -\infty$ when approaching  from the positive $w$ side. This feature is absent without tilt. In the region where the fixed points are stable, i.e. when
\be{stable}
w_-^2 < w <0\,,
\ee 
we can have  $1 < \epsilon <3/2$. For the eigenvalues and curves $w_i^2$, see  Appendix \ref{Aev2}.     

The full stability of the $w=0$ solutions can not be solved using the simplest Lyaponov method. The linear method showed  
 the fixed point to be stable on this curve in the shear and the vorticity directions. Using the center manifold theorem, we find  the lowest order approximation in the dust direction  to be
\be{dust1}
 \Omega' (\alpha)=-\frac{2 \text{sech}^2(2 \lambda )}{\text{sech}(2 \lambda )+1} \Omega^2 (\alpha)\,.
\ee
From this equation can be deduced that the fixed point is unstable in the dust direction: if the system is perturbed in the positive dust direction, it will return to the fixed point, but if  the system is perturbed in the negative dust direction, it will continue in the negative direction. Consequently, when $w=0$, the system has no stable fixed points.

\subsubsection{The fixed point allowing stable inflation with shear-hair}

\begin{figure}[h]
\centering
\includegraphics[scale=1]{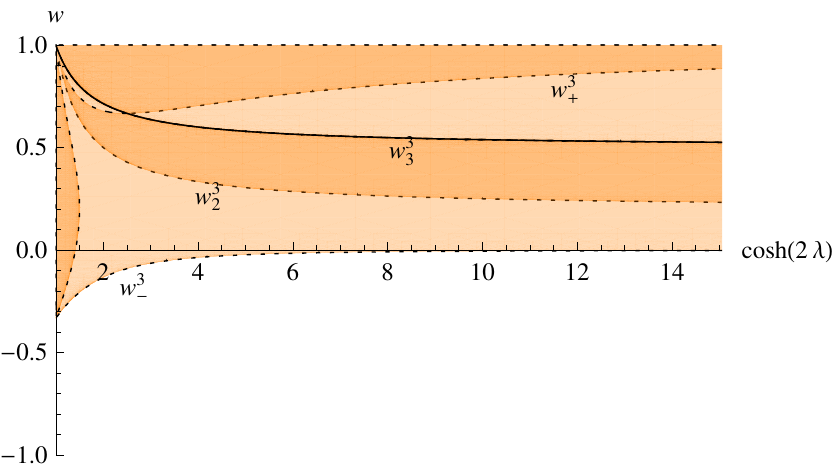} 
\includegraphics[scale=0.8]{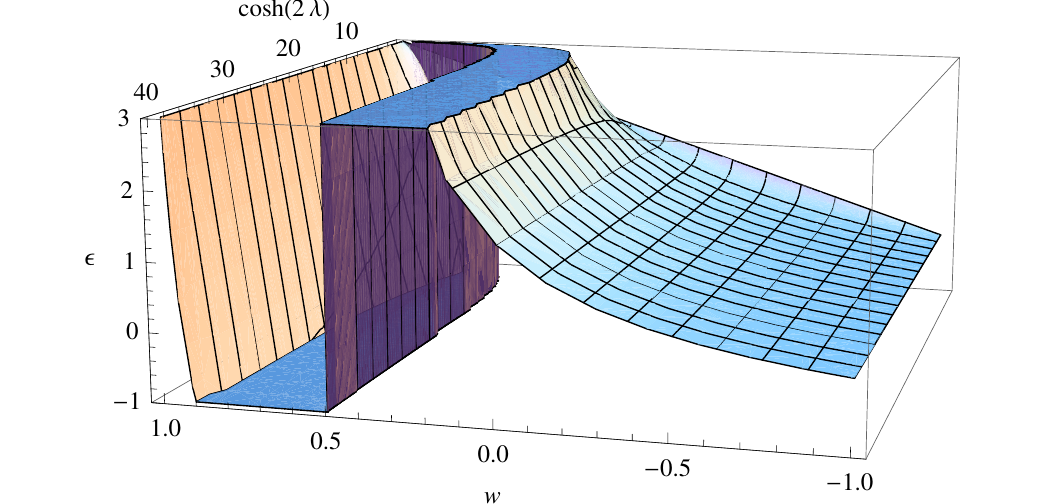} 
\caption{\footnotesize The left panel shows the stability properties of the fixed point (\ref{fp9a}). In the white region the fixed point is an attractor, and a saddle point in the orange regions (repelling in one direction in the lighter and two directions in the darker shaded areas). On the black curves the fixed point is not defined, and on the dotted curves (including $w=0$) a Lyaponov exponent has a vanishing real part. If $\lambda \rightarrow \infty$, then $w_+^3 \rightarrow 1$, $w_3^3 \rightarrow 1/2$, $w_2^3 \rightarrow 1/5$ and $w_-^3 \rightarrow 0$. The right panel shows the values of $\epsilon$. Colors available online.}
\label{kfp98a}
\end{figure}

The nontrivial fixed point where the  vorticity and the  curvature disappear is at
\bea{fp9a}
\V^*&=&0 \nonumber \\
K^*&=&0 \\
\Omega^*&=&\frac{3 (w-1) (3 w-4 W+1)}{W (-3 w+2 W+1)^2} \nonumber \\
\Sigma^*&=&\frac{2-2 W}{-3 w+2 W+1}  \nonumber
\eea
  This situation is depicted in Figure \ref{kfp98a}. 
 The fixed point can represent any expansion rate as it exists for $-\infty<\epsilon<\infty$. The curve where the fixed points are not defined is  $w=(1 + 2 \cosh(2\lambda))/(-1 + 4 \cosh(2\lambda))$, moreover  $\epsilon \rightarrow \infty$ when approaching the curve from the negative $w$ side and $\epsilon \rightarrow -\infty$ when approaching  from the positive $w$ side. This feature is absent without tilt. There exists a wide region of parameter space where the fixed point is stable,  i.e. when
\be{stable2}
 w < w_-^3\,.
\ee
 The eigenvalues and $w_-^3$ are presented in  Appendix \ref{Aev3}. Furthermore, the right panel of Figure \ref{kfp98a} shows  $\epsilon \approx 0$ when $w\approx-1$.  Therefore, this fixed point gives a counter example to the cosmic no-hair conjecture. 

\subsubsection{The shear universe fixed points}

The system includes two fixed points where only the shear do not vanish:
\be{fp3}
\V^*=K^*=\Omega^*=0 \quad \text{and} \quad \Sigma^*=\pm1.
\ee
The eigenvalues of the negative dimensionless shear  fixed point are 
\be{fp3evpos}
\left(6, 6,\frac{3-3 w}{W}\right),
\ee
and the eigenvalues of the positive dimensionless shear   fixed point are 
\be{fp3evneg}
\left(-2, 2,\frac{-3 w+4 W-1}{W}\right).
\ee
Both fixed points are always saddle and have $\epsilon=3$ for all $w$ and $\lambda$.

\subsubsection{The curvature and shear universe fixed points}

The system includes another fixed point independent of the tilt angle and the equation of state parameter: 
\be{fp5}
\V^*=\Omega^*=0 \quad \text{and} \quad K^*=3/4 \quad \text{and} \quad \Sigma^* =1/2.
\ee
The eigenvalues of the  fixed point are 
\be{fp3evpos}
\left(-\frac{3}{2}, -\frac{3}{2},-\frac{3 w}{W}\right).
\ee
These are all negative if $w \in (0,1]$, thus describing a stable fixed point. The solution is not an attractor if the fluid dilutes slower than dust, $w \in [-1,0)$, since then relative contribution of the fluid energy density will dominate in the end. In the case where $w=0$, the stability can not be determined using the linear method, but as in this case this fixed point coincides with (\ref{fp9b}), we already know it is unstable. This solution expands with the rate given by $\epsilon=3/2$, regardless of $w$ and $\lambda$.

\section{Cosmological applicability}
\label{cosmo}

In this section we consider the possible cosmological relevance of the models. First we will discuss the role of dynamical variables of the system (\ref{dqs}), and then construct explicit examples of early and late cosmologies with nontrivial but potentially viable anisotropic dynamics.


\subsection{On the anisotropies}

Here we briefly discuss various possible effects that seem to emerge in these anisotropic models. In particular, we report the observations that 1) the shear-free condition could be supported with the tilt 2) the rotation might grow in special cases 3) strong tilt can have a freezing effect on the evolution and 4) we obtain anisotropically accelerating solutions. The last two of these we illustrate with some numerical solutions obtained by integrating our full equations motions. The numerical solutions are in complete agreement with the analytic considerations of the previous section.

\subsubsection{Shear}

Cosmological observations constrain the shear to be very small in the post-recombination universe. In particular, the CMB is compatible with at most $|\Sigma | \lesssim 10^{-4}$ at the decoupling \cite{Ade:2015lrj}.
This appears to be problematic if we want to have otherwise sizable anisotropies, since according to Eq.\ (\ref{S}), curvature as well as vorticity act as sources for the shear. Thus, generically in viable cosmologies we expect to have only small effects from each of these terms, since it is otherwise difficult to keep the shear under control. We will present in the following some examples with nontrivial but small cosmological effects from anisotropies in the presence of shear. We have found both stable and unstable fixed points with nonvanishing shear, of which (\ref{fp9a}) can accommodate accelerating expansion and can thus describe anisotropic inflation.

With special fine-tuning however, one might eliminate the shear from the anisotropic fixed points  (\ref{fp10}) and  (\ref{fp9b}) (the shear is zero in the fixed point  (\ref{fp9a}) iff $\lambda=0$ or $w=-1$). The fine-tuning required is the specific relation between the fluid parameters $\cosh{2\lambda} = -(1+1/w)/2$ that singles out a zero-measure family of models in the $w<0$ half of the parameter space. We will not investigate these special cases further here, but mention that it might be interesting to consider them as a possible realisation of the so called shear-free condition in cosmology, see Refs. \cite{Mimoso:1993ym,Koivisto:2010dr,Pereira:2015pxa} on theoretical and phenomenological aspects of shear-free cosmologies.




\subsubsection{Rotation}

The vorticity determines an infinitesimal rotation in an infinitesimal proper time interval \cite{Ehlers:1993gf}. Its role differs from the shear and curvature anisotropies, as the vorticity decouples from the expansion. By this we mean that it does not enter the constraints (\ref{F},\ref{e}) and thus does not act as an effective energy source for the expansion, but rather affects the dynamics of the overall scale factor indirectly via its dynamical impact on the evolution of the curvature and the shear. Furthermore, the redshift of a distant object is not directly dependent on $\V$ either, because it does not affect on the separation between two successive spatial hyper surfaces \cite{Ehlers:1993gf}. The  fixed point where the vorticity is not zero, (\ref{fp10}), is always saddle and can represent the shear, the curvature, the radiation or the matter dominated epochs according to possible $\epsilon$ values.

From the evolution equation (\ref{V}) we see that the condition for the vorticity to grow is $\epsilon>1+4\Sigma$. Thus, in a shear-free case we cannot increase the rotation in an accelerating universe. Negative shear could however catalyse the growth of vorticity.



\subsubsection{Tilt}\label{Largetilt}

\begin{figure}[h]
\centering
\includegraphics[scale=0.8]{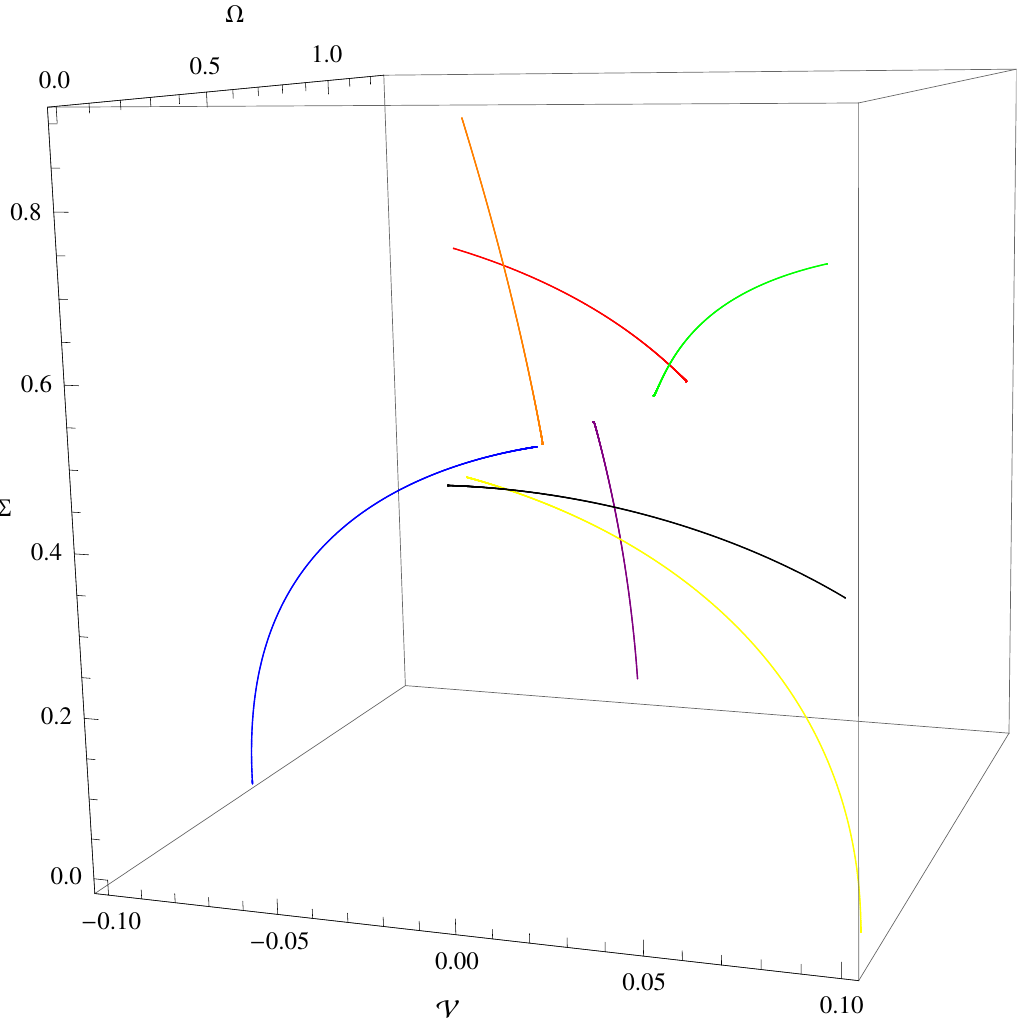} 
\includegraphics[scale=0.8]{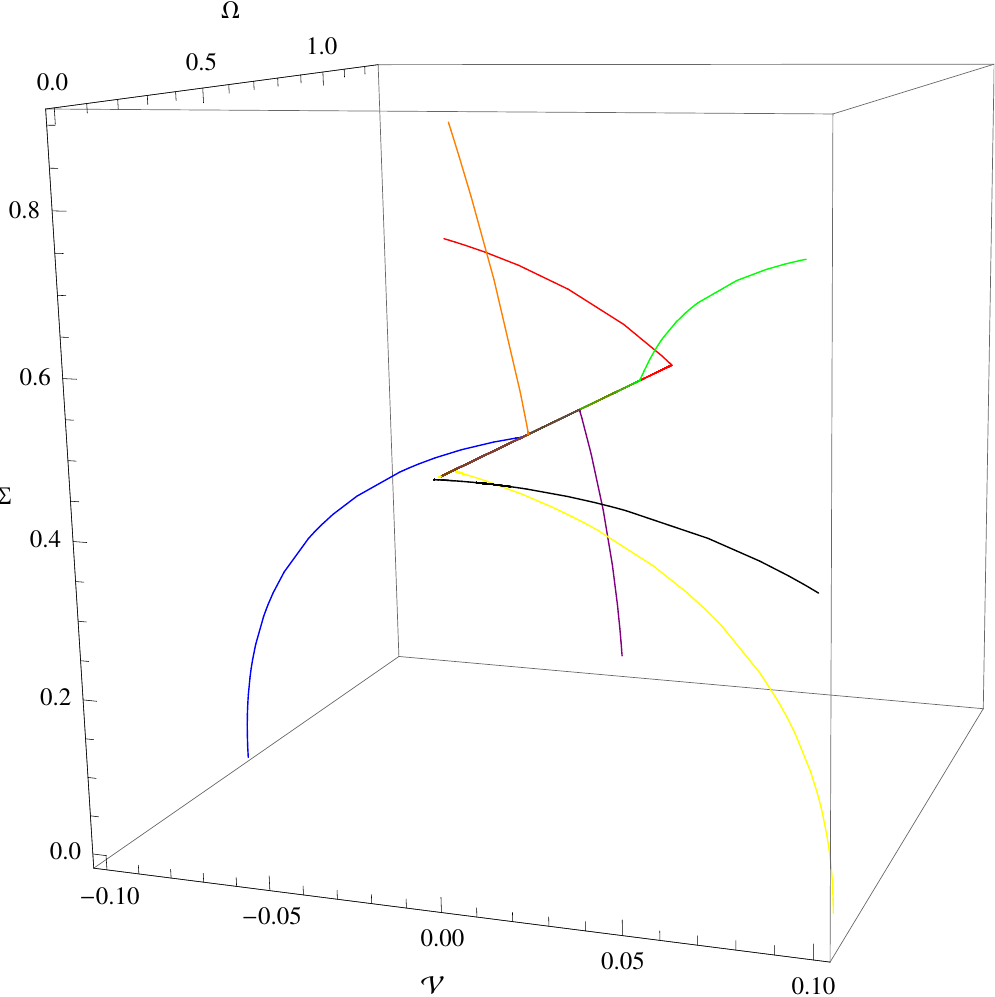} 
\caption{\footnotesize    Trajectories in a dust-filled universe with $\lambda=3$. In the left panel, the system has evolved for for 1000 e-folds  starting from different initial values, but none of the trajectories seemingly has evolved after  5 e-folds. In the right panel, the same trajectories have had additional 200000 e-folds time to evolve, in which time all of them have nearly reached the fixed point (\ref{fp5}). Hence, after 5 e-folds the trajectories do evolve slowly along the same line   parallel to the $\Omega$-axis, where  $\V= 0$ and $\Sigma = 0.5$. See Section \ref{Largetilt} for further discussion. }
\label{dust_k1}
\end{figure}

For strongly tilted models, $\cosh(2 \lambda) \gg1$, the time passing along the trajectories of the system may slow down. Let us illustrate this ''freezing'' effect with numerical solutions. Choosing a positive-density matter with the equation of state $w=0$, we experimentally find that the fixed point (\ref{fp5}) appears to attract a wide range of solutions. The rate at which the solutions converge to this attractor however can depend considerably upon the tilt angle $\lambda$ of the matter fluid. This is seen in Fig. (\ref{dust_k1}), where trajectories  approaching to the fixed point (\ref{fp5}) are drawn with appreciable tilting angle. 
We set the tilt to $\lambda=3$ and evolve models with various initial conditions, all the curves appear to stop evolving after 5 e-folds, where $\V\approx 0$ and $\Sigma \approx 0.5$ (the left panel). However, allowing the integration to continue much longer, say 200,000 e-folds longer,  we notice that all the curves do still approach the fixed point (\ref{fp5}) along the line  $\V= 0$ and $\Sigma = 0.5$ (the right panel). 
In contrast, with say $\lambda=0.1$, the system is close to  the fixed point (\ref{fp5}) already in tens of e-folds. One of the dynamical effects of the tilt is thus obviously that it slows down or "freezes" the evolution (at least in some corners of the phase space near the fixed points).  

\subsubsection{Inflation} \label{inflation}

\begin{figure}[h]
\centering
\includegraphics[scale=1]{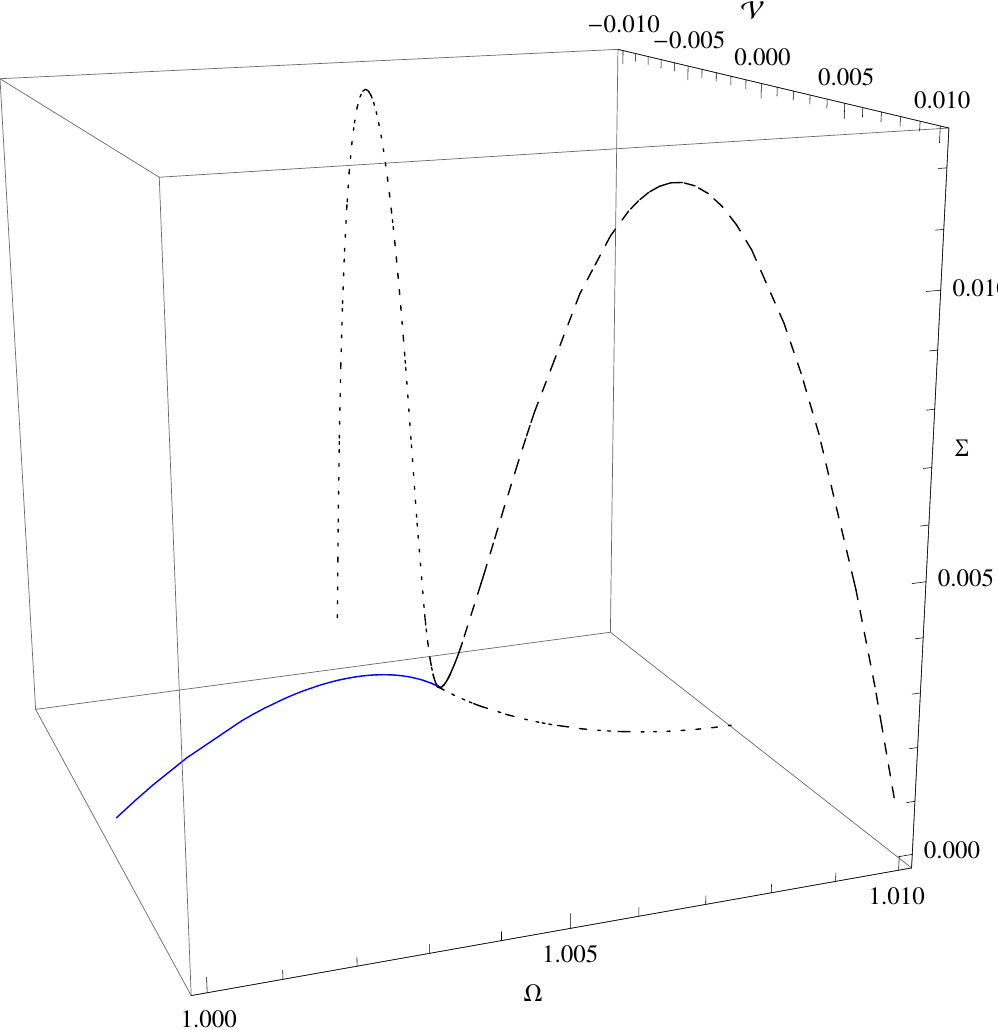}  
\caption{\footnotesize  Four solutions of the system, where $w=-0.99$ and $\lambda =4$,  are drawn in the vicinity of the stable fixed point (\ref{fp9b}). There the system is in exponentially accelerating state, $\epsilon \approx 0.005$, where $\V^*=0$, $\Sigma^* \approx 1.7 \times 10^{-3}$ and $\Omega^* \approx 1.005$, hence presenting a counterexample to the cosmic no-hair conjecture. The solid curve starts from where an isotropic inflation would occur, but it also evolves to the anisotropic fixed point. }
\label{de_k1}
\end{figure}

The fixed point (\ref{fp9b}) in the system can represent a universe in a stable exponentially expanding  phase ($\epsilon\approx 0$). The shear is non-zero at the fixed point, unless $\lambda=0$ or $w=-1$. If $w$ is slightly larger than negative unity, the universe is approximately in an inflationary phase regardless of the tilt. However, a non-zero tilt makes the expansion faster and induces shear. Hence, this fixed point represents a counter example to the cosmic no-hair conjecture. As expected, this requires violationg of the strong energy condition (namely, $w< -1/3$, see Appendix \ref{energyconditions}). The amount of shear is connected to the rate of expansion in this point, and is thus not arbitrary if we want to have inflation. In the limit of large tilt, these quantities are given solely by the equation of state of the fluid:
\begin{equation}
\Sigma^* \rightarrow \frac{1+w}{2-4w}\,, \quad \epsilon^* \rightarrow \frac{3\left( 1+w \right)}{2-4w}\,, \quad \text{when } \lambda \rightarrow \infty\,.
\end{equation}
These limiting values are reached with good accuracy with $\lambda$ of order few. Remarkably, we can have both the shear and the expansion rate within the desired ranges. Namely,
the near-de Sitter inflation with $\epsilon \lesssim 0.05$ that is required to produce the observed tilt of the scalar fluctuation spectrum \cite{Ade:2015lrj} is perfectly compatible with a shear at the level of few percents, which is the maximal amount allowed by constraints on the anisotropy \cite{Ade:2013vbw} (in fact, with large enough tilt simply $\Sigma^*=\epsilon/3$).

Numerical examples of cosmologies converging to the anisotropic inflationary attractor are shown in Fig. \ref{de_k1}, with the models parameters set as $w=-0.99$ and $\lambda =4$. Trajectories are shown with four qualitatively different initial conditions, including an isotropic case corresponding to standard inflationary expansion. We see that, from all the initial conditions, the universe ends up in the inflationary fixed point that efficiently brakes the rotation but retains a finite shear supported by the tilt. 
 


\subsection{Bianchi IX cosmology with late-time anisotropies} \label{cosmological model}

We will construct here an explicit example of full cosmological evolution. The purpose is to find a concrete scenario wherein the cosmological parameters fit the observations and the anisotropies are not too large to be immediately ruled out by the CMB bounds, but wherein we could still generate a small amount of rotation (that could have relevance to the large-angle CMB anomalies). We consider a scenario where dark energy is tilted. The equations to integrate are Eqs. (\ref{S}-\ref{F}), but in order to describe full realistic cosmic history, we need  include also  dust and radiation in the play and supplement the corresponding terms in these equations. 
At low redshifts we then expect that our tilted dark energy component will begin to dominate the energy budget (over dust and radiation) and in addition to accelerating the universe, introduce slight anisotropic features. The aim of the numerical study is to investigate the detailed dynamics within the latter transition period as well as  at the time of decoupling.

 In this set-up then dust and radiation are comoving with the coordinates and the dark energy fluid is tilted with respect to them. 
To distinguish dust, radiation, and dark energy from each other, we can still treat the variables (\ref{dqs}) as effective ones and merely decompose the energy-momentum tensor (\ref{Tab2}) as
\be{emt}
T^{ab}=(\rho_2+p_2)n^an^b+p_2g^{ab}+(\rho_3+p_3)n^an^b+p_3g^{ab}+(\rho_1+p_1)u_1^au_1^b+p_1g^{ab}+q_1^a u_1^b+u_1^aq_1^b,
\ee 
where  $n^a=(1,0,0,0)$ is the four velocity comoving with the coordinates and
\be{u1}
u_1^a=\left(\cosh (\lambda_1 ),0,0,\sinh (\lambda_1 ) e^{2 \beta (t)-\alpha (t)}\right) 
\ee
 is the four velocity of the tilted fluid and $\lambda_1$ is the exact tilt of the tilted fluid with respect to the comoving coordinates. The details of the decomposition are given in Appendix \ref{decomposition}.  

 By using the energy-momentum tensor (\ref{emt}), the relevant field equations become
\begin{eqnarray} \label{e2}
\epsilon &=& \frac{1}{2}  \left[ 3w_3  -1  \right] \Omega_{ 3}(\alpha)+ \frac{1}{2}  \left[ 3w_2  -1  \right] \Omega_{2}(\alpha)+ \frac{1}{2}  \left[ 3w_1  -1  \right] \Omega_1(\alpha)-K(\alpha) +2+\Sigma^2(\alpha)\,,   \qquad \qquad
\end{eqnarray}
\begin{equation}\label{F2}
1=\Omega_{3}(\alpha)+\Omega_{2}(\alpha)+ W_1 \Omega_1(\alpha)+
K(\alpha)+ \Sigma^2(\alpha)\,, \qquad \qquad  \qquad \qquad \qquad \qquad \qquad \qquad \qquad \quad \, \,
\end{equation}
when it comes to the two constraint equations, and the four dynamical equations of motion generalise to
\begin{eqnarray} \label{S2}
\Sigma'(\alpha) = -1+2
  K(\alpha)+\Sigma(\alpha){}^2+(\epsilon-3)
  \Sigma(\alpha)+  \V(\alpha){}^2+ \Omega_1 (\alpha)+ \Omega _{2 }(\alpha)+ \Omega _{3 }(\alpha)\,, \qquad \qquad \qquad \qquad 
\end{eqnarray}
\begin{eqnarray}\label{O2}
\Omega_1 '(\alpha)
&=&\Omega_1(\alpha) \left(2 \epsilon +2 \Sigma (\alpha)-4 +\frac{1-3 w_1-2 \Sigma (\alpha)}{W_1}\right) \qquad \qquad \qquad \qquad \qquad \qquad \qquad \qquad \qquad \quad  \quad \,\,\,\,\,  \nonumber \\
&&-\frac{\Omega' _{2 }(\alpha)+\Omega _{2 }(\alpha)[3(1+w_2)-2\epsilon]+\Omega _{3 }'(\alpha)+\Omega _{3 }(\alpha)[3(1+w_3)-2\epsilon]}{W_1}\,,
\end{eqnarray}
\begin{equation}\label{V2}
\V'(\alpha)=
  \left(\epsilon-4 \Sigma(\alpha)-1\right) \V(\alpha)\,,\qquad \qquad \qquad \qquad \qquad \qquad \qquad \qquad \qquad \qquad \qquad \qquad \qquad \qquad \qquad 
\end{equation}
\begin{equation}\label{K2}
K'(\alpha)=-2\left\{K(\alpha)[-\epsilon+1+\Sigma(\alpha)]+\Sigma(\alpha) \V(\alpha)^2\right\}\,,\qquad \qquad \qquad \qquad \qquad \qquad \qquad \qquad \qquad \qquad \quad \,\,
\end{equation}
where $W_1 \equiv (-w_1+1+\frac{w_1+1}{ \text{cosh}(2 \lambda_1 )})/2$, $\Omega_i=\kappa \rho_i/(3H^2)$ and $w_i=p_i/\rho_i$ ($i=\{1,2,3\}$).  
Like in Section \ref{model}, the conservation of energy-momentum tensor yields only one equation,  even though the decomposition of $T_{\mu \nu}$ results two new functions. To obtain a closed system, we are required to impose two new equations.
The geometrical mean of the expansion is $e^{\alpha}$, hence imposing $\rho_i \propto e^{-3(1+w_i)\alpha}$  we find in terms of the dimensionless variables 
\be{Od}
\Omega_{i}'(\alpha)=\Omega_{i}(\alpha)[2 \epsilon-3(1+w_i)]
\ee
 Using this relation for $\Omega_2$ and $\Omega_3$, Eq. (\ref{O2}) simplifies to
\begin{eqnarray}\label{O3}
\Omega_1'(\alpha)
&=&\Omega_1(\alpha) \left(2 \epsilon +2 \Sigma (\alpha)-4 +\frac{1-3 w-2 \Sigma (\alpha)}{W_1}\right)\,.
\end{eqnarray}

After choosing appropriate values for the parameters $w_i$ and  $\lambda_1$, Eqs. (\ref{e2}), (\ref{F2}), (\ref{S2}), (\ref{V2}), (\ref{K2}), (\ref{Od}) and (\ref{O3}) constitute a full autonomous system which we use to describe the evolution of the universe from the radiation dominated epoch to the present time. Eventhough earlier we discussed about a set up where dark energy is tilted with respect to dust and radiation, the aforementioned equations are derived without assuming which fluid is tilted. Hence in the subsequent, we give special attention to tilted dark energy scenarios, but  employ  other possibilities too.

\subsubsection{Numerical solution}

We studied the above specified system of equations  (\ref{e2}), (\ref{F2}), (\ref{S2}), (\ref{V2}), (\ref{K2}),  (\ref{Od}) and (\ref{O3})  numerically in order to quantify the amount of fine-tuning required for the system to describe viable late time cosmology, including the CMB anomalies. In practice we mean by viability now that with anisotropies kept at the maximum of some percent level, the system should evolve through radiation and matter dominated eras as usual, and then near the present epoch approach a dark energy dominated solution so that  $\Omega_{\Lambda}\approx 0.7$ and $\Omega_d\approx 0.3$ and $\Omega_r   \sim 10^{-5}$ today. The time from the decoupling to the present time then corresponds to $\ln(1100) \approx 7$ e-folds (with shear at the maximum percent level, the expansion is direction-independent to a good approximation). From Eq. (\ref{kq1}) it appears that the tilt shouldn't be too strong in order to avoid generating too much shear in the expansion. We thus impose $\lambda\leq1/100$, and furthermore choose the fluids to be dark energy, dust and radiation. 

According to our numerical analysis, the solutions satisfying the above described conditions all originate from the vicinity of the negative or positive fixed point (\ref{fp3}). However, not all the initial conditions close to the fixed points (\ref{fp3}) lead to the desired type of solutions, but the initial conditions need to be carefully chosen. From the vicinity of either of the fixed points  (\ref{fp3}), the solutions enter the radiation dominated era alike, but can lead to different late time cosmology. Nevertheless, constraining  $K$ and $\V$ from above appears to be sufficient to keep the anisotropies negligible, independently  of the sign of $\Sigma$,  which fluid is tilted or the amount of tilt (as long as $\lambda\leq1/100$). These results indicate that inflation is not required to vaporise shear  for viable late time cosmology to occur. Let us consider  examples where some shear is left after inflation.

\begin{figure}[h]
\centering
\includegraphics[scale=1]{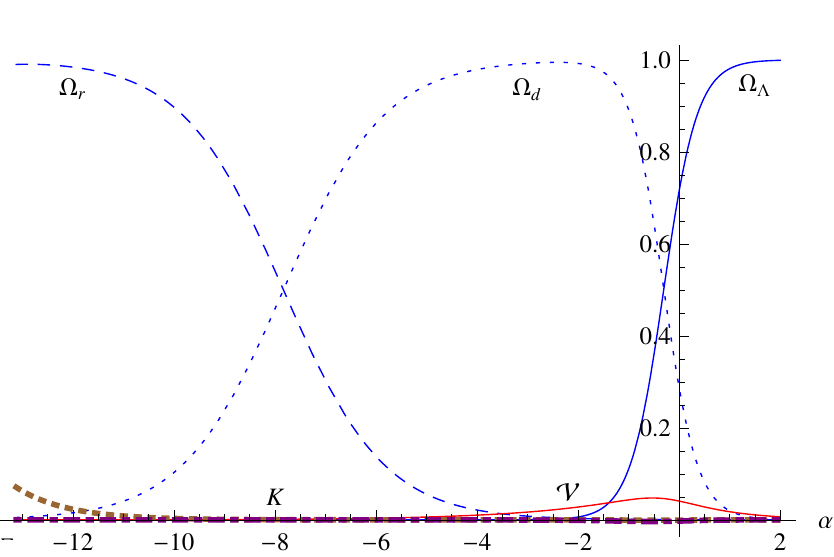} \,\, \,\,
\includegraphics[scale=1]{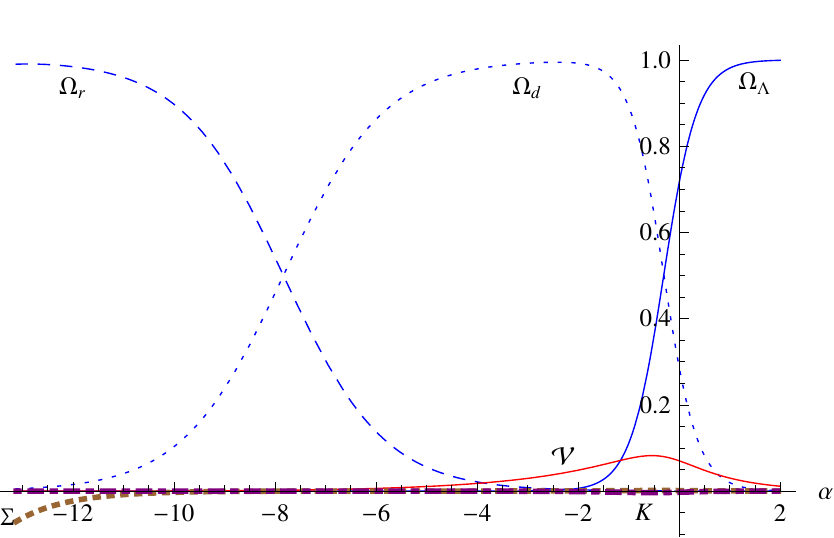} \, \,
\includegraphics[scale=1.]{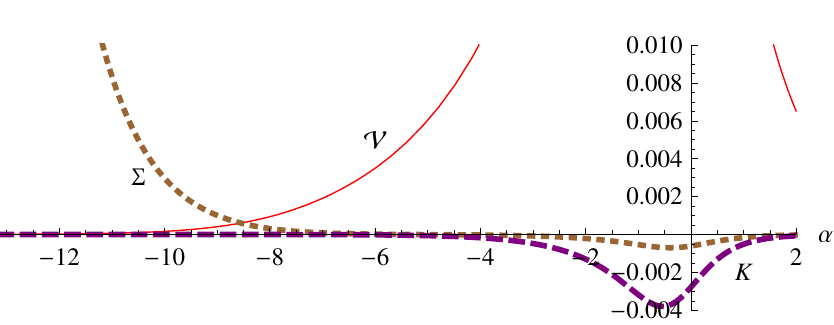} \,\, \,\,
\includegraphics[scale=1.]{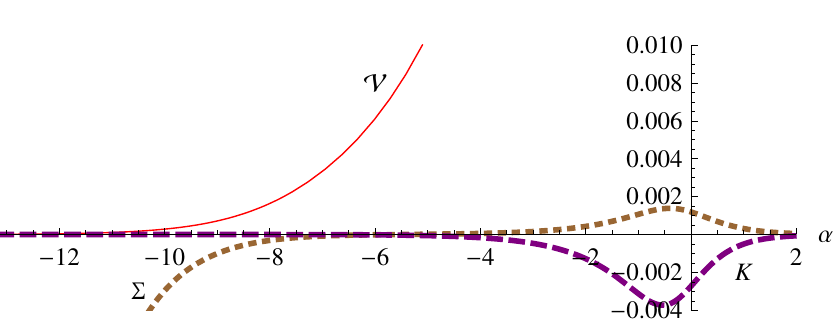} \quad \quad
\caption{\footnotesize Bianchi IX cosmological models, Example A on the left side and Example B on the right side, describing the evolution of the universe from radiation dominated epoch to the dark energy dominated epoch. The system is solved using Eqs. (\ref{e2}), (\ref{F2}), (\ref{S2}), (\ref{V2}), (\ref{K2}), (\ref{O3}) and (\ref{Od}) and the initial conditions in Table 1 .  The horizontal axis is the number of e-folds from the present time. All the evolving quantities are plotted in the  upper panels, whereas the behavior of  $\V$, $\Sigma$, and $K$ is magnified close to the horizontal axis in the lower panels.}
\label{cosmology}
\end{figure}

{\bf Example A:} The tilted fluid is dark energy and $\lambda=1/100$.  The initial conditions  are given in Table \ref{ics} and the solution is drawn in Fig. (\ref{cosmology}), showing the dynamical variables as functions of the number of e-folds ($\alpha$). In addition, the values of the dynamical quantities at the decoupling and at the present time are tabulated in Table \ref{ics}. We observe the emergence of three epochs: the radiation, the dust and the dark energy dominated expansions, the latter finally converging to the attractor (\ref{fp9b}) as expected. Also the shear, the vorticity and the curvature are drawn in the figure, though they are almost indistinguishable. They reach the peak of their dynamical significance around the matter - dark energy equality, where we roughly have $\Sigma \approx -0.0007$,  $K \approx -0.004$, and  $\V \approx 0.05$. Eventually these will decay with the accelerated expansion, since also shear given by the asymptotic future solution (\ref{fp9b}) vanishes when $w=-1$.

 Let us study study this example further. According to our numerical simulations, at the beginning of the radiation epoch only conditions $\abs{K}\lsim 10^{-10}$ and $\abs{\V}\lsim 10^{-5}$ are to be satisfied to ensure small enough anisotropies during late time cosmology. These conditions are remarkably alike to those stipulated for the standard $\Lambda$CDM model referred as the flatness problem. Breaking the condition for $K$ would make either $\abs{\Sigma}$ too large at the decoupling or both $\abs{\Sigma}$ and $\abs{K}$ too large during the transition from dust dominated to dark energy dominated eras. Also the vorticity is constrained from above, since too much rotation would boost the $\Sigma$ and $K$ during the transition period. In addition to these initial conditions, of course, we need to choose the relative amounts of the three types of energy densities, $\Omega_i$'s, suitably to have proper radiation, dust and dark energy epochs. The initial condition for $\Sigma$ is then to be determinate using Eq.\ (\ref{F2}).  Further numerical  analysis indicates that, at the accuracy presented in Table \ref{ics} and Figure \ref{cosmology}, the results of Example A are the not dependent on the size of $\lambda$ (assuming $\lambda\lsim1/100$) and hold for both signs of  $\V$ and irrespectively which fluid is tilted.

{\bf Example B:} Let us consider the same scenario as in Example A, but change the sign of the initial shear.  The initial conditions   are given in Table \ref{ics} and the solution is drawn in Fig. (\ref{cosmology}). In addition, the values of the dynamical quantities at the decoupling and at the present time are tabulated in Table \ref{ics}. The system appears identical to that in Example A apart from $\Sigma$ and $\V$. From the figure can be seen that the ratio $|\V/\Sigma|$  takes the higher value at $\alpha \sim -0.5$ in the case of negative initial shear and at the same $\alpha$ the shears take opposite signs. From Table 1 we see that these conclusions hold at the decoupling and present time too.  Further numerical  analysis indicates that the same constraints as in Example A, namely $\abs{K}\lsim 10^{-10}$ and $\abs{\V}\lsim 10^{-5}$, apply here for the same reasons. Moreover,  at the accuracy presented in Table \ref{ics} and Figure \ref{cosmology}, the system in this example does not alter by changing the fluid that is tilted,  the size of  $\lambda (\lsim1/100$) or the sign of $\V$.

\begin{figure}[h]
\centering
\includegraphics[scale=1]{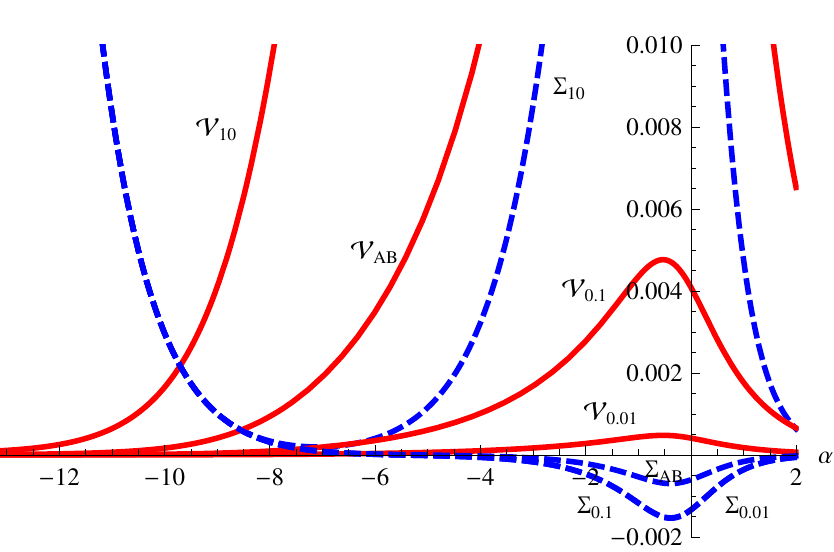} \,\, \,\,
\includegraphics[scale=1]{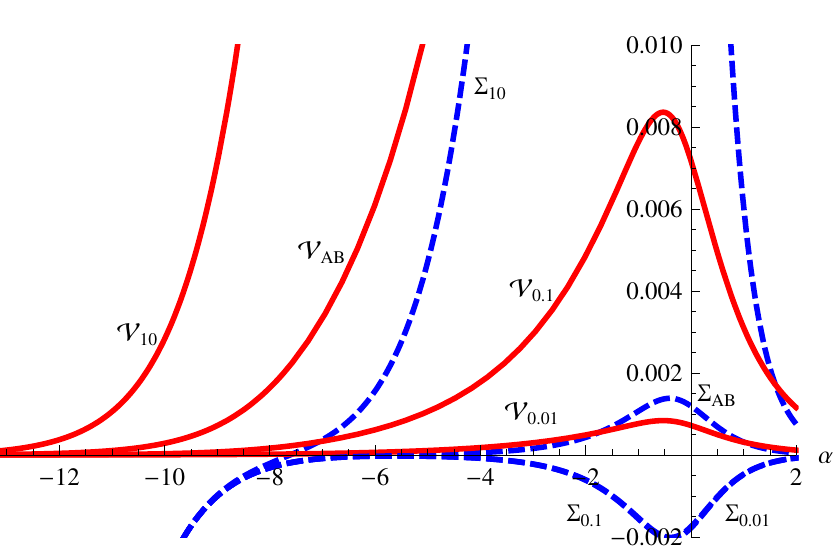} \, \,
\caption{\footnotesize The evolution of $\Sigma$ and $\V$  in two Bianchi IX cosmological models is integrated using several  initial values for vorticity, where the initial shear is positive (left plot) and negative (right plot). Curves denoted by $\V_i$ and $\Sigma_i$ represents the evolutions of the vorticity and the shear, respectively, and the subscript refers to the initial value of the vorticity (see Example 3).  The curves $\Sigma_{0.01}$ and  $\Sigma_{0.1}$ are indistinguishable  from each other in both panels.}
\label{fe3}
\end{figure}

{\bf Example C:} Let us study Examples A and B using different values for initial vorticity. We  impose the  initial values
 $$(0.01 \times  \V_{AB},\, 0.1 \times  \V_{AB},\,  \V_{AB},\,  10 \times  \V_{AB}),$$
 where $\V_{AB}$ denotes the initial value for $\V$ used in Examples A and B, i.e. $\V_{AB}=1 \times 10^{-5}$,  and represent the results in Table \ref{te3} and in Figure \ref{fe3}. Furthermore, the evolution of $\Sigma$ and $\V$ corresponding to the initial value $X \times  \V_{AB}$ is denoted by $\Sigma_X$ and $\V_X$ in Figure \ref{fe3}, respectively. For example, the curves  $\Sigma_{0.01}$ and $\V_{0.01}$  describe the evolution of  $\Sigma$ and $\V$ corresponding the initial vorticity $0.01 \times  \V_{AB}$.  The effects of changing the initial $\V$ appear alike in both cases. Curves $\Sigma_{0.01}$ and  $\Sigma_{0.1}$ are indistinguishable for both cases in Figure \ref{fe3}, indicating   initial $\V\lsim 0.1 \V_{AB}$ has negligible effect on the evolution of $\Sigma$. Increasing initial vorticity further  induces positive contribution to shear and vorticity in both cases around the time of dust - dark energy equality. Qualitatively the behavior is similar but quantitatively weaker at the decoupling and at the present time (see Table \ref{te3}). Increasing the initial vorticity  $\V \lsim \V_{AB}$  by an order of magnitude induces an order of magnitude growth for $\V$ but negligible growth for $\Sigma$ at the decoupling. The results are independent on the amount of tilt, assuming $\lambda\lsim 1/100$, and which fluid is tilted.

The peculiar feature that all the results exhibit independence on the size of  $\lambda\lsim 1/100$ and which fluid is tilted  imply that the evolution of the anisotropies is decoupled from the fluid for small enough tilt.\footnote{To support this conclusion, we explored the range $1/100<\lambda$  to a lesser extent and found  the anisotropies  begin to grow during the domination of the tilted fluid. The only exception was tilted dark energy, because $w=-1$ makes $\lambda$ disappear from the dynamical equations.} In the corner of the phase space studied above, the anisotropies  reach their peak values during matter - dark energy transition. This is no surprise,  because after that point they can no longer grow as they approach the  asymptotic future solution (\ref{fp9b}) cutting all hair of anisotropy when $w=-1$, hence this location is at least a local maximum of anisotropies.   Because also shear and vorticity appear to be decoupled  for a small enough tilt and vorticity at the beginning of the radiation era, the results obtained here differ significantly form another Bianchi type IX realisation \cite{Barrow1985}, where even small vorticity was found to induce considerable shear. In Ref.  \cite{Barrow1985} the authors found the maximum value  for the present time vorticity  to be $\sim 3.9 \times10^{-13}$ for non-axisymmetric Bianchi IX metric sourced by tilted perfect fluid including only dust. Around 10 magnitude difference in the results suggests that different realisations of the Bianchi IX space-time can yield remarkably different results. In the light of Ref. \cite{Barrow1985} and our numerical examination, the model used here shows potential explaining the anomalies in the CMB and late time expansion simultaneously.

\begin{table} \caption{The initial values for the dynamical variables of the models in Examples A and B are given,  the exact value of shear  is to be calculated from the generalised Friedmann equation (\ref{F2}), but only its approximative value is presented in this table. The systems are solved using these initial data and the values of the dynamical variables at the decoupling time and  at the present time are reported with one digit accuracy.} \label{ics} 
\begin{center}
  \begin{tabular}{| c | c | c | c | c | c | c |} 
    \hline 
            &   $\OL$ &  $\Od$ & $\Or$ & $\Sigma$ & $\V$ & $K$\\ \hline
\multicolumn{7}{|c|}{Example A - positive initial shear}  \\ \hline
  Initially & $1 \times 10^{-19}$ & $5 \times 10^{-3}$ & 0.99 & $\approx 7.07107 \times 10^{-2}$ &$1 \times 10^{-5}$ & $-1 \times10^{-10}$ \\ \hline
 At the Decoupling $\sim$ & $1 \times 10^{-9}$ & 0.7 & 0.3 & $9 \times 10^{-5}$ & $2 \times 10^{-3}$ &  $-6 \times 10^{-6}$ \\ \hline
Present Time $\sim$ & 0.7 & 0.3 & $1 \times 10^{-4}$ & $-6 \times 10^{-4}$ & $4 \times 10^{-2}$ & $-3 \times 10^{-3}$ \\    \hline
\multicolumn{7}{|c|}{Example B -  negative initial shear}  \\ \hline
  Initially & $1 \times 10^{-19}$ & $5 \times 10^{-3}$ & 0.99 & $\approx -7.16378 \times 10^{-2}$ &$1 \times 10^{-5}$ & $-1 \times10^{-10}$ \\ \hline
 At the Decoupling $\sim$ & $1 \times 10^{-9}$ & 0.7 & 0.3 & $-8 \times 10^{-5}$ & $3 \times 10^{-3}$ &  $-6 \times 10^{-6}$ \\ \hline
Present Time $\sim$ & 0.7 & 0.3 & $1 \times 10^{-4}$ & $1 \times 10^{-3}$ & $7 \times 10^{-2}$ & $-3 \times 10^{-3}$ \\    \hline
  \end{tabular} 
\end{center} 
\end{table}

\begin{table} \caption{ The values of $\V$ and $\Sigma$ at the decoupling and at the present  time corresponding to different initial values for vorticity are presented. Here,  $0.01 \times  \V_{AB}$,  $0.1 \times  \V_{AB}$, $\V_{AB}$, and $10 \times  \V_{AB}$ represent different initial vorticities (see Example 3) and on the left side the values correspond to positive initial shear (see Example A) and on the right side the values correspond to negative initial shear (see Example B). } \label{te3}  
\begin{center}
  \begin{tabular}{| c | c | c | c | c | c| c | c | c | c |  } 
    \hline 
            &   $0.01\times  \V_{AB}$ &  $0.1 \times  \V_{AB}$ & $\V_{AB}$ & $10 \times  \V_{AB}$   & &  $0.01 \times  \V_{AB}$ &  $0.1 \times  \V_{AB}$ & $\V_{AB}$ & $10 \times \V_{AB}$ \\ \hline
&\multicolumn{4}{|c|}{At the decoupling}&  & \multicolumn{4}{|c|}{At the decoupling}  \\ \hline
  $\V$ & $3 \times 10^{-5}$ & $3 \times 10^{-4}$ &  $3 \times 10^{-3}$ &$3 \times 10^{-2}$ & & $2 \times 10^{-5}$ & $2 \times 10^{-4}$ &  $2 \times 10^{-3}$ &$2 \times 10^{-2}$ \\ \hline
 $\Sigma$ & $-9 \times 10^{-5}$ & $-9 \times 10^{-5}$ & $-9 \times 10^{-5}$ &  $3 \times 10^{-4}$&  & $8 \times 10^{-5}$ & $8 \times 10^{-5}$ & $9 \times 10^{-5}$ &   $2 \times 10^{-4}$ \\ \hline
&\multicolumn{4}{|c|}{At the present time} & & \multicolumn{4}{|c|}{At the present time}  \\ \hline
 $\V$ & $7 \times 10^{-4}$ & $7 \times 10^{-3}$ &  $7 \times 10^{-2}$ &$3 \times 10^{-1}$&& $4 \times 10^{-4}$ & $4 \times 10^{-3}$ &  $4 \times 10^{-2}$ &$3 \times 10^{-1}$ \\ \hline
 $\Sigma$ & $-2 \times 10^{-3}$ & $-2 \times 10^{-3}$ & $-1 \times 10^{-3}$ &  $4 \times 10^{-1}$ & & $-1 \times 10^{-3}$ & $-1 \times 10^{-3}$ & $-6 \times 10^{-4}$ &  $3 \times 10^{-2}$ \\ \hline
   \end{tabular} 
\end{center} 
\end{table}

\section{Conclusions}
\label{conclu}

We investigated the dynamics of the tilted axisymmetric Bianchi IX model with fluid allowing energy flux. The cosmological application we had in mind was the generalisation of anisotropic inflation and anisotropic dark energy models to generate parity-violating imperfect features at cosmological scales. The latter could perhaps then be manifest in the large-angle anomalies of the CMB. In the framework of the Bianchi IX model, accommodating a tilt of the fluid required us to introduce a heat flux for consistency. In general, the Einstein equations and the continuity equation are given treating the equation of state $w$ and the tilt angle of the fluid $\lambda$ as time dependent functions, enabling us to consider realistic multi-fluid scenarios, but for simplicity the phase space was studied by reducing $w$ and $\lambda$ to parameters.

We performed a dynamical system analysis, identifying seven fixed points in the cosmological phase space and determining their stability for given model parameters $w$ and $\lambda$,  which we considered in the ranges $-1\leq w \leq 1$ and $-\infty<\lambda<\infty$. Only one of the fixed points allows for rotation, but always describes non-accelerating expansion. The fixed points generically include a non-negligible amount of shear, which can then be accompanied by the curvature with or without the tilted fluid energy density. Of particular interest for possible applications is an inflating attractor in which the amount of shear is given by the tilt of the fluid. The vorticity in general has a different status in the system compared with the other dynamical variables, as the vorticity does not contribute directly to the expansion rate.

The anisotropically accelerating fixed point can be relevant to both early inflation and  dark energy models. We did some preliminary numerical investigations of both types of scenarios, verifying that indeed realistic cosmological evolutions can be constructed. The late time solutions show potential explaining both the parity preserving and violating  anomalies of the CMB, which is unexpected in the light of earlier studies of the subject \cite{Barrow1985}. Our numerical studies indicate that  the only fine-tuning the late time solutions require is to set the curvature very close to zero after inflation.  The fixed point (\ref{fp9b}) retains a constant amount of shear (when $w\sim -1$ and $\lambda \neq 0$) with the expansion, even if it is accelerating, thus providing a counter example to the cosmic no-hair conjecture. Though many such counter examples have been presented (see the Introduction \ref{introduction} for an incomplete list of references), such have not been, previously to our knowledge, deviced by exploiting tilt and energy flux. 

It could be very interesting to study further the observational implications of such possible imperfections in the source fluid. Determining the precise signatures from relaxing the standard assumptions $q^a=0$ and $\lambda=0$ could allow us to falsify the proposed origin of cosmic anomalies as a tilted dark energy (or inflaton). However, before such proposal could be promoted from the present phenomenological exploration into a more convincing alternative cosmology, we needed to develop also the theoretical underpinnings of the model. It would be desirable to describe the fluid by a lagrangian field theory, instead of adding it as a parameterised energy component by hand. In particular, in the present context the precise form of the heat flux was fixed by mathematical consistency rather than physical motivations. To describe the dynamics of fluid more properly, we would guess one probably needs to allow in general time-dependent $w(\alpha)$ and $\lambda(\alpha)$, and the dynamics for these should be determined from the lagrangian principle for the fundamental degrees of freedom for the fluid.

Despite these challenges, we believe our results could provide an interesting starting point for further explorations of rotating universes filled by some kind of tilted dark energy.

\acknowledgments

We would like to thank Jos\'e Pedro Mimoso for useful discussions on anisotropic cosmologies and comments on the draft.
This study is partially (P. S.) supported by the Magnus Ehrnroothin S\2\2ti\3 foundation (6.3.2014) and Turun Yliopistos\2\2ti\3 foundation, identification number 11706.

\bibliography{refs}

\appendix

\section{Energy conditions} \label{energyconditions}

The energy-momentum tensor (\ref{Tab2}) does not violate the weak or dominant energy conditions. For some parameter values $w$ and $\lambda$, it does however, break the strong energy condition. The strong energy condition for (\ref{Tab2}) reads, using the results of Ref. \cite{Kolassis1988} for our case,
\be{sec}
\rho  \left(-\sqrt{(w+1)^2 \tanh ^2(2 \lambda )}+w+1\right) \geq 0 \qquad \rho  \left(\sqrt{(w+1)^2 \text{sech}^2(2 \lambda )}+2 w\right) \geq 0\,.
\ee
For $w\geq 0$, the above conditions are satisfied for any $\lambda$, but the latter of the conditions is violated by any $\lambda$  if  $w<-1/3$. Thus neither the tilting of the fluid nor the implied energy flux changes the usual bound for the violation of the strong energy condition.

\section{Eigenvalues} \label{Aev}

\subsection{The rotating fixed point (\ref{fp10})}
\label{Aev1}

The eigenvalues $\ev_i^1$ of the stability matrix of the fixed points (\ref{fp10}) are
\be{evfp1}
\ev_1^1=3\,\frac{3 w+2 W-1}{5 W-1}, \qquad \ev_2^1=3\,\frac{ \sqrt{W}( w- W)+\sqrt{\mathcal{A}}/2}{\sqrt{W}(5
   W-1)}, \qquad \ev_3^1=3\,\frac{ \sqrt{W}( w- W)-\sqrt{\mathcal{A}}/2}{\sqrt{W}(5
   W-1)}\,,
\ee
where $\mathcal{A}=-2 w \left(3 w^2+w-3\right)+8 (w+14) W^2+40 (w-1) w W-92 W^3-28 W+2$. The dotted curves separating the different areas in Figure \ref{kfp10} are 
\be{kp1c}
w^1_{\pm}=\frac{\cosh ^2(2 \lambda )-2 \left(1 \pm\sqrt{3-2 \cosh (2 \lambda
   )}\right) \cosh (2 \lambda )+2}{\cosh (2 \lambda ) (\cosh (2
   \lambda )+4)-2}\,.
\ee
Curves $w^1_{\pm}$ satisfy the equality in   (\ref{fp10c}) and
 equations $\ev_2=K^*=\V^*=0$.

\subsection{The non-rotating matter-scaling fixed point (\ref{fp9b})}
\label{Aev2}

The eigenvalues $\ev_i^2$ of the stability matrix of the fixed points (\ref{fp9b}) are
\be{evfp2}
\ev_1^2=-\frac{3}{2}\,\frac{3 w+2 W-1}{2 W-1}, \qquad \ev_2^2=-\frac{3}{4}\,\frac{W  (w-2 W+1)-\sqrt{\mathcal{B}}}{ (1-2 W) W}, \qquad \ev_3^2=-\frac{3}{4}\,\frac{ \left(W  (w-2 W+1)+\sqrt{\mathcal{B}}\right)}{ (1-2 W) W}\,,
\ee
where  $\mathcal{B}=W \left(-24 w^3+w^2 (W+16)+w \left(28 W^2-30 W+8\right)+(1-2 W)^2 W\right)$. The curves separating the different areas in Figure \ref{kfp98b} are obtained from  equations $\ev_i=0$ (dotted) and $1/\ev_i=0$ (solid), $i=\{1,2,3\}$. These curves are
\bea{kp2c1}
w^2_1=\frac{-1}{2 \cosh (2 \lambda )+1}, \quad w^2_2=0, \quad  w^2_3=\frac{1}{\cosh (2 \lambda )-1},  \quad w^2_4=\frac{\cosh (2
   \lambda )+1}{\cosh (2 \lambda )-1}\,, \\ \label{kp2c2}
w^2_{\pm}=\frac{1-\cosh (2 \lambda )+\cosh ^2(2 \lambda )\pm \cosh (2
   \lambda ) \sqrt{\cosh ^2(2 \lambda )-2 \cosh (2 \lambda
   )+5}}{2 \cosh ^2(2 \lambda )+2 \cosh (2 \lambda )-1} \,,\, \, \, \, 
\eea
and it is easy to verify that they approach to zero or unity as given in Figure \ref{kfp98b}.
The $+$ case of Eq.\ (\ref{kp2c2}) yields the lower bound for inequality (\ref{stable}).

\subsection{The non-rotating flat matter-scaling fixed point (\ref{fp9a})}
\label{Aev3}

The eigenvalues $\ev_i$ of the stability matrix of the fixed points (\ref{fp9a}) are
\be{evfp3}
\ev_1=-\frac{3 (w-1) (3 w-4 W+1)}{2 W (-3 w+2 W+1)}, \quad \ev_2=\frac{3 \left(-3 w^2+2
   w+(1-2 W)^2\right)}{W (-3 w+2 W+1)}, \quad \ev_3=\frac{-9 w^2+6 w+24 W^2-24 W+3}{-6 w W+4
   W^2+2 W}\,.
\ee
The curves separating the different areas in Figure \ref{kfp98a} are obtained from  equations $\ev_i=0$ (dotted) and $1/\ev_i=0$ (solid), $i=\{1,2,3\}$. These curves are
\bea{kp3c1}
w_1^3=1, \quad w_2^3=\frac{\cosh(2\lambda)+2}{5 \cosh(2\lambda)-2}, \quad  w_{\pm}^3=\frac{1-\cosh(2\lambda)+\cosh^2(2\lambda) \pm \cosh(2\lambda) \sqrt{\cosh^2(2\lambda)-2\cosh(2\lambda)+5}}{2
   \cosh^2(2\lambda)+2 \cosh(2\lambda)-1}\,, \\
\tilde{w}_{\pm}^3=\frac{2-2\cosh(2\lambda)+\cosh^2(2\lambda)\pm 2\cosh(2\lambda)\sqrt{3-2 \cosh(2\lambda)}}{\cosh^2(2\lambda)+4 \cosh(2\lambda)-2},  \quad w_3^3=\frac{2 \cosh(2\lambda)+1}{4\cosh(2\lambda)-1}\,. \qquad \qquad \, \,
\eea
At the limit $\lambda \rightarrow \infty$, the behaviour of  curves $w_{\pm}^3$, $w_2^3$, and $w_3^3$ is given in Figure \ref{kfp98a}.

\section{Decomposition of the energy-momentum tensor} \label{decomposition}

In the set of Eqs. (\ref{S})-(\ref{F}),  the fluid variable $\Omega$ can be considered as an effective ''total'' fluid, whose density consists of a combination of separate components $\Omega_i$, as $\Omega=\Sigma_i \Omega_i$. The effective or the total equation of state parameter $w$ describes how the density of the total fluid is related to its pressure and the effective tilt illustrates how much the total fluid is tilted with respect to the normals of the surfaces of homogeneity. 
In this sense, the quantities $\Sigma$, $K$, $\V$, and $q^a$ are also to be considered as effective. 
Therefore, we can decompose $\Omega$ and $q^a$  into the "original" components, but still treat the other quantities in their effective forms.

Consider two four-velocities, $u_1^c=(\cosh(\lambda_1),0,0,\sinh(\lambda_1)/\sqrt{g_{33}})$ and  $n^c=(1,0,0,0)$. The average (or effective) four-velocity, $u^c$, is the linear combination  $a u_1^c+bn^c$, where $a$ and $b$ are constants   so that $u^cu_c=-1$. By introducing a shifted angle $\lambda$ implicitly via
$$
a=\frac{\sinh(\lambda) }{\sinh(\lambda_1)}\,, \quad \text{and}   \quad
b=\cosh(\lambda)-\frac{\sinh(\lambda) }{\sinh(\lambda_1)}\cosh(\lambda_1)\,,
$$
the new four-velocity is nothing but $u^c=(\cosh(\lambda),0,0,\sinh(\lambda)/\sqrt{g_{33}})$. Hence, the tilt angle $\lambda$ can be interpreted as an average (or effective) tilt of the combination of the fluids with respect to the reference frame.

The relations between the effective and exact fluid densities, state parameters and tilt angles can now be obtained by equating tensors (\ref{Tab2}) and (\ref{emt}). However, a simpler expression is achieved by expressing (\ref{emt}) using $W_1$, 
\be{emtuus2}
T=\left(
\begin{array}{cccc}
 -W_1(t) \rho_1(t) - \displaystyle\sum_{i=2}^3 w_i(t) \rho_i (t)  & 0 & 0 & 0 \\
 0 & \displaystyle\sum_{i=1}^3 w_i(t) \rho_i (t) & 0 & 0 \\
 0 & 0 &  \displaystyle\sum_{i=1}^3 w_i(t) \rho_i (t)  & \cos (\theta ) (W_1(t)-1) \rho_1 (t) \\
 0 & 0 & 0 & [w_1(t)+W_1(t)-1] \rho_1 (t)+\displaystyle\sum_{i=2}^3 w_i(t) \rho_i (t) 
\end{array}
\right),
\ee
and  equating (\ref{emtuus}) and (\ref{emtuus2}), yielding
\bea{dec}
\rho(t) = \rho _1(t)+\rho _2(t)+\rho_3(t), \,\, w(t)= \frac{w_1(t) \rho _1(t)+w_2(t) \rho _2(t)+w_3(t)\rho_3(t)}{ \rho_1(t)+\rho_2(t)+\rho_3(t)},  \,\,
W(t)=\frac{W_1(t) \rho_1(t)+\rho_2(t)+\rho_3(t)}{ \rho_1(t)+\rho_2(t)+\rho_3(t)}. \,\,\,
\eea
Substituting these expressions to Eqs. (\ref{S}) - (\ref{F})  yields Eqs.  (\ref{e2}) - (\ref{K2}) if  $w_i$ and $\lambda_1$ are constants.

\end{document}